\documentclass[12pt]{iopart}
\usepackage{iopams}  
\usepackage{graphicx}\usepackage{epsfig}\usepackage{portland}
\usepackage{bm}
\usepackage{epic}\usepackage{eepic}
\usepackage{exscale}\usepackage{rotating}
\usepackage{dcolumn}    
\usepackage{color}
\usepackage{ulem}
\def\FepGeV{$^{56}$Fe$_{(1\,A\,\textrm{GeV})}+p$ }

\def\beam{{\mathrm{p}}}
\def\Lab{{\mathrm{L}}}

\def\diff{\mathrm{d}}
\def\si{\sigma_{\mathrm{I}}}
\def\vvec{\boldsymbol v}
\def\uvec{\boldsymbol u}
\def\pvec{\boldsymbol p}

\def\vpar{v_{\|}}
\def\vper{v_{\bot}}
\def\vvpar{\vvec_{\|}}

\def\vperm{\widetilde{v}_{\bot}}
\def\vvperm{\widetilde{\vvec}_{\bot}}

\def\ddVu#1{\displaystyle{\diff #1\over\diff\vvec^u}}

\def\dyield{\diff I(v_{\|}^{\beam})}

\def\yield{\displaystyle{\dyield\over\diff\vpar^{\beam}}}

\def\aver#1{\langle #1 \rangle}
\def\deltausigmaI{\delta_u \sigma_{\mathrm{I}}}
\begin{document}

\title[Inclusive selection of intermediate-mass-fragment formation modes]{
	Inclusive selection of intermediate-mass-fragment\\
	formation modes in the spallation of $^{136}$Xe
}

%
%
\author{P.Napolitani$^{1,2}$, K.-H.Schmidt$^{2}$ and L.Tassan-Got$^{1}$}
\address{$^{1}$~IPN Orsay, Universit\'e Paris-Sud 11, CNRS/IN2P3, 91406 Orsay cedex, France}
\address{$^{2}$~GSI, Planckstr. 1, 64291 Darmstadt, Germany}
\ead{napolita@ipno.in2p3.fr}

%
%
\begin{abstract}
	A correlation between the production and kinematic 
properties of the fragments issued of fission and 
multifragmentation is established in the study of the reaction 
$^{136}$Xe~+~hydrogen at 1 GeV per nucleon, measured in inverse 
kinematics at the FRagment Separator (GSI, Darmstadt).
	Such observables are analysed in a comprehensive study, 
selected as a function of the decay mode, and related to the 
isotopic properties of the fragments in the intermediate-mass 
region.
	Valuable information can be deduced on the characteristics 
of the heaviest product in the reaction, which has been considered 
a fundamental observable for tagging the thermodynamic properties 
of finite nuclear systems.
\end{abstract}

%
%
\pacs{25.40.Sc, 25.70.Pq, 24.60.Dr, 29.30.Aj, 29.87.+g}

%
%
\vspace{2pc}
\noindent{\it Keywords}: 
Longitudinal-velocity spectra, isotopic cross sections, intermediate-mass fragments, inclusive isotopic identification

\maketitle
%
%
%
%
\section{
    Introduction				\label{section1}
}
%
%
	We present a comprehensive experimental study of the decay 
processes occurring 
in the spallation of $^{136}$Xe induced by 1 GeV protons, 
through inclusive kinematic and production observables.
	The incident energy, the size and the neutron enrichment of
the system we chose is compatible with the threshold for 
multifragmentation.
	This situation is expected to be characterised by fluctuations 
between asymmetric fission and multifragmentation.
	Such scenario has already been presented in a previous work~\cite{Napolitani07},
where cross sections were published for the whole production 
distribution, without any selection on the reaction mechanism.
	In this work
we develop a selection method to identify 
and quantify the relative shares of the decay processes, fission
and multifragmentation, which lead to the production of 
intermediate-mass fragments. 
	Once established a correlation between production and 
kinematic quantities as a function of the isotopic observable and 
the decay mode, we analysed the characteristics of the 
intermediate-mass fragments issued of fission and 
multifragmentation, with the further aim of determining a set of 
observables 
relevant for describing the decay phenomenology.

\subsection{Context}
    In relativistic nuclear reactions, especially when induced by 
hadrons or light nuclei, kinematic and thermal observables probe
the transition between different reaction mechanisms as a function 
of the bombarding energy~\cite{Hufner84,Lynch87}.
    Already some decades ago, several experiments were focused on 
spallation with the purpose of studying the reaction 
mechanism (for a non-exhaustive series of pioneering experiments see 
refs.~\cite{Grover62,Kaufman76,Warwick82,Hirsch84,Andronenko86,Barz86,Kotov95,Hsi97,Avdeyev98}).
	In nucleon-induced reactions, 
at the incident energy of few hundred MeV per nucleon, an 
excited and fully equilibrated compound nucleus~\cite{Sanders99} is formed, which 
decays by the emission of fragments in the form of particle 
evaporation and fission, according to the number of final states 
available for different mass splits~\cite{Moretto75,Moretto89}.
	At the incident energy of several GeV per nucleon, the highly 
excited composite nuclear system has also a large probability of 
decaying by multifragmentation.
	For a comprehensive review on this subject see ref.~\cite{Cole2000};
see also ref.~\cite{WCI} for a large collection of reviews on this process, 
and refs.~\cite{ISIS,Karnaukhov99,Karnaukhov03} for specific 
reviews treating multifragmentation induced by high-energy protons.
	In nucleon-induced reactions at incident energies which are 
intermediate between these extreme situations, for instance around 
1 GeV per nucleon, the multifragmentation threshold could be 
attained: in this case, the exit channel of the system is determined 
by the competition between multifragmentation and sequential decay.

	Such a complex phenomenology inspired the formulation of 
several alternative descriptions, attempted since the early 
eighties, in order to understand the transition between the 
different reaction mechanisms. 
	Initially, a description of multifragmentation was proposed
as a process of successive binary splits~\cite{Moretto88} within 
the transition-state model, or as an extension of the saddle-point 
interpretation to highly excited systems~\cite{Lopez94}.
	Later, several experimental data gave arguments for the
large success of statistical models, 
and descriptions based on the phenomenology of phase 
transitions~\cite{Bondorf95,Raduta97,Gross01}, 
so that binary splits might even appear 
among the possible fragment configurations.
	Not only light-ion induced reactions comply largely with 
the statistical nature of the process (ISIS experiments~\cite{ISIS}), but also recent 
heavy-ion experiments at Fermi energy 
(INDRA~\cite{Tabacaru03,Lopez06,Pichon06,Bonnet07,Bonnet07b,Bonnet08}, 
MULTICS~\cite{Bruno08}) gave special prominence to such a picture.
	As a function of the violence of the collision, the bimodal 
distribution of the largest-fragment size is interpreted as a 
signal of the finite-size counterpart of the liquid-gas phase 
transition in nuclear matter~\cite{Gulminelli02, Chomaz02}, which is 
associated to a finite latent heat~\cite{Gulminelli03}: the energy (and 
density) of the system and the size of the largest fragments are 
order parameters which rule the evolution of the system from a 
configuration where a large cluster survives at low excitation 
energy to a configuration where the system disassembles in small 
fragments of comparable size at high-excitation~\cite{Lopez05, Gulminelli07}.
	It should be pointed out that a dedicated study of the 
effect of the Coulomb field on the fragmentation process further 
enriched this picture and revived the interest on the phenomenology
of binary fission in highly excited systems.
	In particular, also the effect of Coulomb frustration applies:
this is a rather general phenomenon originally introduced in condensed-matter 
studies to describe a large variety of systems, ranging from
magnets on specific lattices to liquid crystals, from spin
glasses to protein folding. 
	The common trait of these systems is the absence of a state where all 
interaction energies are minimised simultaneously.
	In the nuclear fragmentation process, Coulomb frustration describes
the situation when the long-range repulsive Coulomb field and the 
short-range attractive nuclear interaction can not be minimised simultaneously,
and results in one additional phase transition process, from the liquid 
phase to fission.
	This phenomenon of Liquid-Fission transition is not 
associated to any latent heat.
	It was observed and detailed within the Liquid Gas 
Model~\cite{Lehaut09, Gulminelli09} as well as in phenomenological 
nuclear statistical models~\cite{Gulminelli03,Chaudhuri09}.

\subsection{Purpose of this work}
%
	The experimental study detailed therein analyses the 
kinematic, production and isotopic observables of the two decay 
modes, fission and multifragmentation, tagged by the production of 
a specific nuclide in the intermediate-mass region (located below
half the mass of the projectile, and ranging down to lithium isotopes).
	In the case of fission, such nuclide is the lightest binary 
partner, the other being close to the mass of the projectile.
	In the case of multifragmentation, such nuclide is one of 
the several fragments of similar size produced by the division of 
the source and has only a fraction of the mass of the projectile.
	In both cases, indirectly or more directly, the 
experimental signature we obtain by the kinematics of the 
intermediate-mass fragments gives a good information on the size of 
the heaviest product in the reaction.
	Especially in the case of multifragmentation, the 
characteristics of the heaviest fragment have a fundamental role in
carrying robust signals of the finite-size counterpart of the 
nuclear liquid-gas phase transition.
	From such premises, the main aim of this experimental work 
is threefold:
\begin{enumerate}
\item
Analysis.
We focus on the selection problematics for high-resolution inclusive experiments, and we propose a new analysis procedure for correlating production and kinematic observables.
\item
Experimental results.
We provide several inclusive intermediate-mass fragment
observables, as a function of the decay process: we quantify both
isotopic cross sections and invariant cross sections, concerning
the production and the kinematics, respectively.
\item
Discussion.
Concerning the exit channel:
we analyse the boost imparted to the partners of binary
splits, we extract the distribution of fragments along the mass
number, and we analyse the strong staggering which characterises the
nuclide production.
Concerning the connection between the entrance channel and the exit channel:
from the kinematic observable, we propose a new inclusive indicator for the impact parameter and we illustrate a possible interpretation for the evolution of the kinematic observable as a function of the size of the fragments for the different decay modes.
\end{enumerate}
%
\section{
	Experimental overview			\label{section2}
}
%
The present study focuses on a situation where the
	contribution of two processes to the decay,
fission and multifragmentation, makes experimental conditions 
rather challenging.
	On the other hand, these experimental conditions are greatly 
simplified by the choice of relativistic energies: in the case of
relativistic nuclear reactions induced by hadrons, the fraction of 
the bombarding energy which remains in the system appears almost 
completely as thermal energy, 
reflected in a mostly isotropic kinematics. 
	Under these conditions, an inclusive approach based on the use 
of a magnetic spectrometer is at present a very well adapted solution
for connecting the 
production observables to the kinematic observables: in particular, 
the production observables consist in the full isotopic 
identification, extended to the complete range of fragment sizes;
the kinematic observables are the distribution of longitudinal 
velocities, measured for single nuclides, from the magnetic 
rigidity, with very high precision and without any threshold.
	Nevertheless, we show in the following that some 
difficulties still remain when rather heavy fragments are studied, 
because the analysis procedure we adopted, based on the 
disentangling of the full velocity distribution in more components, 
becomes limited.
%
%
\subsection{Spallation at the multifragmentation threshold} \label{section2A}
    A first inclusive technique of disentangling reaction processes 
on the basis of longitudinal-velocity spectra was introduced for 
distinguishing the average characteristics of the fission fragments 
from those of the evaporation residues~\cite{Enqvist01,Bernas03},
in inverse-kinematic experiments at the FRagment 
Separator~\cite{Geissel92,Schmidt87} at GSI in Darmstadt.
	Concerning spallation at the multifragmentation threshold,
still with the same facility and experimental approach, the 
kinematic and production properties of intermediate-mass fragments 
were deduced in an inclusive experiment for the reaction 
$^{56}$Fe$+$ hydrogen at 1 GeV per 
nucleon~\cite{Napolitani04,Villagrasa07}.
	With the purpose of reconstructing the invariant-velocity 
spectra of intermediate-mass fragments from inclusive kinematic 
observables, a simple procedure of analysis of the 
longitudinal-velocity spectra was introduced: it was based on the 
assumption that all exit channels were associated to the same value 
of the linear momentum transfer for one individual nuclide formed 
as a final product.
	The experimental results revealed mixed signals of 
multifragmentation and fission~\cite{Napolitani04}, which could 
however not be quantified in their relative share and identified 
separately; a possible disentangling of the two exit channels in 
terms of kinematic and thermal properties could 
only be suggested through the comparison with a statistical model.
	The same system was also the object of an exclusive experiment 
at ALADIN~\cite{LeGentil08} at GSI.
	Profiting from particle correlations, but not reinforced by 
kinematic observables, it confirmed that intermediate-mass 
fragments were prevalently produced in binary decays, and also 
revealed the presence of a few decays with larger fragment multiplicity.
	Also in this case, the analysis of the reaction mechanism was 
discussed on the basis of statistical models.
%
%
\begin{figure}[t]\begin{center}
\includegraphics[angle=0, width=0.375\textwidth]{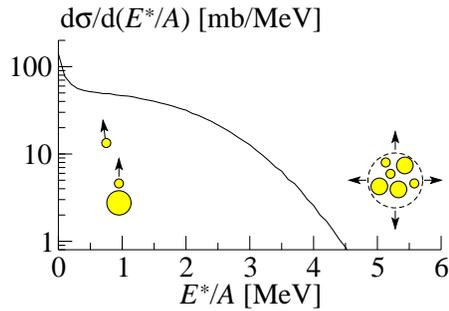}
\end{center}\caption
{
	Intra-nuclear-cascade calculation of the distribution of 
excitation energy of the hot fragments in the system
$^{136}$Xe$_{(1\,A\,\mathrm{GeV})}+p$ (the intranuclear 
exciton-cascade model of ref.~\cite{Gudima83} was used)
Above around 3 MeV per 
nucleon the onset of multifragmentation is expected.
Elsewhere, fission-evaporation is the dominant decay process.
}
\label{fig1}
\end{figure}

    In a more recent measurement at the FRagment Separator, the 
same inclusive experimental approach, previously used for iron, 
was followed to measure the system $^{136}$Xe~+~hydrogen at 1 GeV 
per nucleon~\cite{Napolitani07}.
	$^{136}$Xe is the stable nucleus with the largest neutron 
excess $N-Z$ among the nuclei below the Businaro-Gallone 
point~\cite{Businaro55a,Businaro55b}, where the saddle point 
becomes unstable to asymmetric splits.
	Binary decay should result in the production of neutron-rich 
intermediate-mass fragments.
    A large feeding of evaporation-fission decays should result 
from the range of excitation energies explored by this system.
	To corroborate our expectations, we calculate the excitation-energy 
distribution of the warm system excluding any cooling mechanism.
	As a rather general example for this kind of calculation, we 
employ the intranuclear exciton-cascade model of ref.~\cite{Gudima83}.
	In fig.~\ref{fig1}, the excitation-energy distribution obtained
for the warm system is shown, and it should be remarked that 
its range exceeds 3 MeV per nucleon.
	According to the large amount of theoretical and 
experimental results~\cite{WCI}, the transition towards multifragmentation is 
expected above this value and, similarly to fission, it 
should also manifest in the formation of neutron-rich 
intermediate-mass fragments.
	Since the reaction products from these two different channels feed the same region of the nuclear chart, isotopic cross sections alone are an insufficient 
information for studying the reaction mechanism.
	Two experimental strategies, based on coupling production and kinematics observables, are suited for this purpose.
	One is the exclusive strategy: its state of the art is the coupling of
a $4\pi$-detector with a spectrometer (at Fermi and lower energies: 
see ref.~\cite{Chbihi07}) for probing both  the initial excitation process and
the excitation channel; in this case the isotopic identification is however 
limited to very small charges.
	The second strategy followed in the present work is the inclusive
measurement of high-resolution velocity spectra of the entire distribution of fragments larger than alpha particles, identified in mass and charge;
in this case a mayor limitation is the absence of event-by-event observables
like correlations and the excitation.
\subsection{Inclusive event selection} \label{section2B}
	We detail in this paper the new method of inclusive event 
selection which we developed for the analysis of the system 
$^{136}$Xe$+p$.
	In comparison with the analysis method used for the 
system $^{56}$Fe$+p$, the new method has the advantage of being a 
quantitative technique of event selection, allowing to associate 
the probability for a given decay pattern to a group of events.
	In particular, the inclusive event selection method is used to
identify kinematically the two processes of fission and 
multifragmentation in the formation of intermediate-mass fragments, 
and to quantify them in terms of nuclide cross sections and 
kinematic properties, benefiting from the two main specificities of
the experimental method we employed.
	The first specificity is the possibility to measure small velocities in the 
projectile frame with high resolution, so as to access the physics of the 
small-energy portions of the kinetic-energy spectra of fragments; 
this portion is found in this work to give constraints for 
the identification of the exit channel.
	The second specificity is the possibility to connect such kinematic
observable to the production of a given nuclide.
\subsection{The FRagment Separator} \label{section2C}
%
%
\begin{figure}[t]\begin{center}
\includegraphics[angle=0, width=0.8\textwidth]{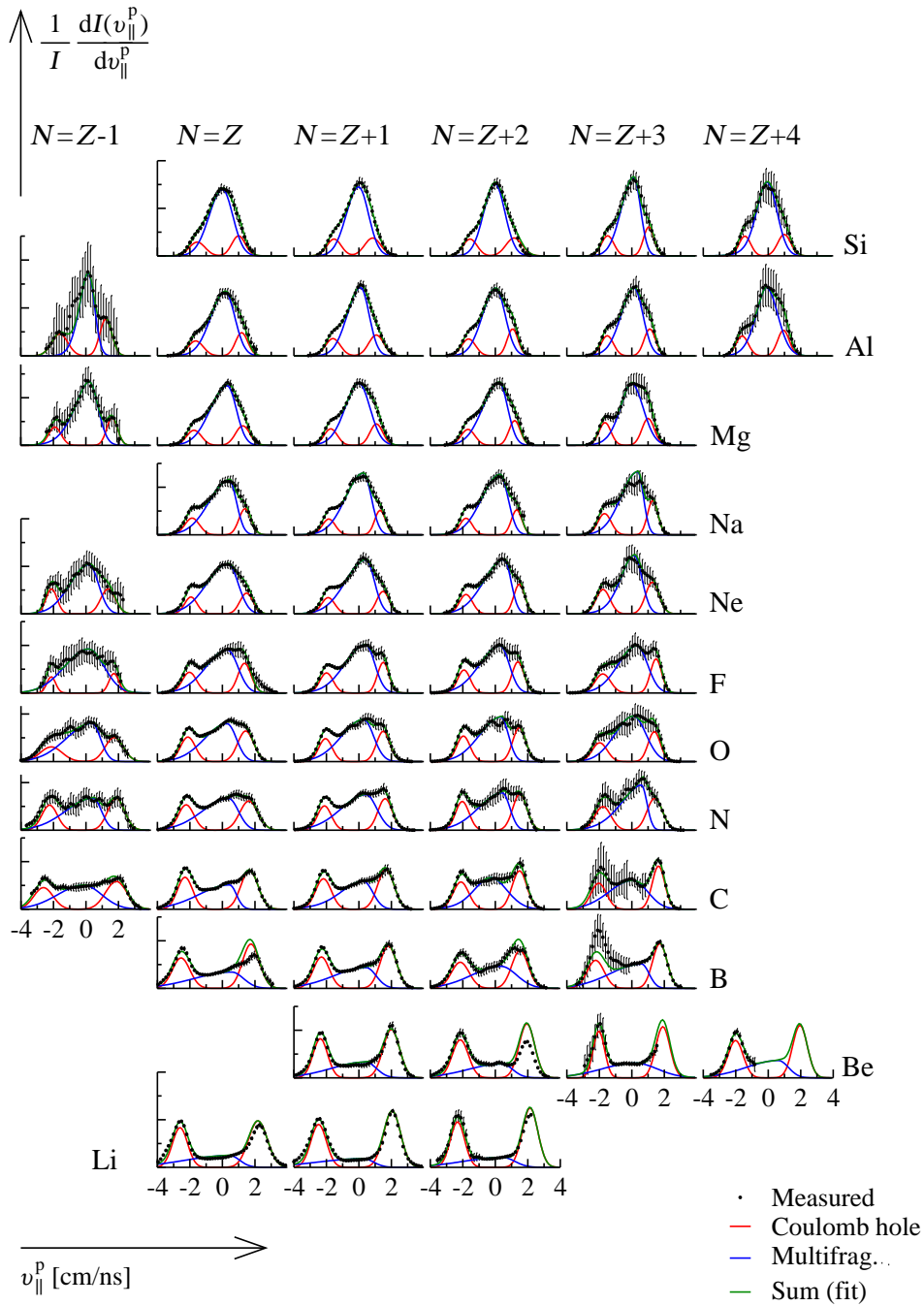}
\end{center}\caption
{
	Systematics of measured velocity distributions 
as a function of the nuclide for elements ranging from Li to Si,
indicated with black dots; error bars show the statistical uncertainties.
	The measured spectra are disentangled by a fitting
procedure  (see section~\ref{section3}) in the contribution of a coulomb-hole mode and a 
multifragmentation mode (see section~\ref{section4}). 
	The shapes of these two components, which are concave and convex,
respectively, are deduced so that their sum is a fit of the measured shape.
}
\label{fig2}
\end{figure}
	The fragments were produced in inverse kinematics by directing 
a primary beam of $^{136}$Xe at 1 $A$ GeV on a target of liquid 
hydrogen contained in a cryostat with thin titanium windows.
	The projectile residues were analysed in-flight, with the
Fragment Separator~\cite{Geissel92,Schmidt87}.
    Four dipole magnets compose the magnetic spectrometer, a 
dispersive focal plane is localised behind the first two dipole 
magnets, and the full spectrometer was set achromatic.
    The identification of a fragment was deduced from the 
measurement of its time of flight, from recording the position where its 
trajectory intersects the dispersive focal plane, and from its nuclear 
charge measured in ionisation chambers, placed in the proximity of the 
terminal focal plane.
	Recoil velocities and kinetic energies were deduced from the 
magnetic rigidities, which were measured by the spectrometer with a 
relative uncertainty of $5\cdot 10^{-4}$ (FWHM) for individual 
reaction products.

	A series of corrections and calibrations were required for 
deducing the distribution of longitudinal recoil velocities for 
each nuclide.
	The raw spectrum was slightly corrected by taking into account 
the energy loss in the target.
	The events related to reactions in any layer of matter which 
does not correspond to hydrogen were suppressed; these events come 
for the greatest part from the material (mostly titanium) 
constituting the cryostat, and their contribution was measured with 
a specific experiment
\footnote{
In particular, the present experimental technique imposed that several 
calibration experiments in identical conditions had to be repeated also with
the empty titanium cryostat in order to exactly subtract any contribution from
non-hydrogen materials~\cite{Napolitani07}
}.
	The number of counts was normalised to the beam dose per 
target thickness, which was deduced by measuring the current
of the primary beam.
	The data treatment required at this point to correct for the decrease of
the angular transmission for ions with magnetic rigidities which deviate from 
the nominal values.
	This effect is seen in the velocity spectra, when they are constructed from 
composing different magnetic settings of the spectrometer: they present 
structures which appear periodically for each portion of the spectrum 
corresponding to one magnetic setting~\cite{Napolitani04}.
	Since the angular transmission can be formulated simply as a function 
of the position of the trajectory intersecting the dispersive focal plane and
the terminal focal plane~\cite{Benlliure02}, these structures are removed by 
correcting the data in a way that they correspond to the transmission of ions 
with nominal magnetic rigidities.
	Finally, additional corrections took into account the 
modification of the yields due to secondary reactions occurring in 
the hydrogen target and in other layers of matter intercepting the 
beam of fragments.
	These corrections are discussed in details in 
ref.~\cite{Napolitani07} and they are the standard procedure of
analysis for all nuclide-production measurements at the FRagment 
Separator.

	At this stage, a distribution of measured events as a 
function of the longitudinal recoil velocity in the projectile 
frame $\vpar^\beam$ is associated to each reaction product, 
identified in mass number and nuclear charge.
	The integral $I$ of each distribution is equal to the fraction 
of production cross section which is selected by the angular 
acceptance of the spectrometer.
	Fig.~\ref{fig2} shows a systematics of measured 
longitudinal-velocity spectra $\diff I(\vpar^\beam)/\diff\vpar^\beam$
normalised to the same integral for all measured isotopes of the 
elements ranging from lithium to silicon.
	In order to obtain invariant-cross-section distributions 
independently of the experimental 
conditions, the geometry of the angular acceptance of the 
spectrometer needs to be accurately accounted for in the data 
analysis.

	In the following section, we describe the method for reconstructing 
the invariant-cross-section distribution 
from the measured longitudinal-velocity spectra.
	Such technique is a further development of the method used 
in the analysis of the spallation of iron at 1 GeV per nucleon 
measured at the FRagment Separator~\cite{Napolitani04}.
%
\section{
    Inclusive velocity spectra	\label{section3}
}
	The FRagment Separator has a small angular acceptance of
$\approx 15$~mr in the laboratory frame.
	This is of course a disadvantage (see discussions in 
refs.~\cite{Napolitani07,Enqvist01,Bernas03,Napolitani04,Benlliure02,Ricciardi06});
on the other hand, even if the angular cuts should be accurately accounted for in
the data analysis, the combination of the reduced acceptance, the achromatic
configuration and the relativistic beams has a special advantage as a
counterpart: no magnetic aberration affects the resolution and all the focal
planes (achromatic and dispersive) are fully defined.
	In particular, only two positions of a fragment trajectory (the
intersections with the dispersive and the achromatic focal planes) are
sufficient for achieving the nominal resolution of $5\cdot 10^{-4}$ (FWHM) for 
the magnetic rigidity.
	For comparison, in a large-acceptance spectrometer, the counterpart of the
broad angular acceptance is a complex form of the magnetic field, which demands
to focus the analysis on an accurate trajectory reconstruction
(as an example see refs.~\cite{Pullanhiotan2008a,Pullanhiotan2008b}).
	The present experiment results in the reversed situation: in a 
small-acceptance achromat operated with relativistic beams, all the effort
should be focused on the treatment of the angular cut by deconvoluting the
momentum spectra of the fragments, but this procedure can be performed with high
precision due to the well defined achromatic ion optics.
	Such precise knowledge of the ion optics is the key of our analysis
procedure: it passes through the complex but conventional deconvolution of the
inclusive velocity spectra which we detail below.
	
Before entering into the analysis, we also introduce some physical
assumptions which will be used.
	The analysis procedure is based on the picture of the reaction 
process proceeding by a fast stage, characterised by nucleon-nucleon
collisions, which ends up in an excited system, and a slower 
deexcitation phase~\cite{Serber47}.
	The interaction with the target proton induces a mean shift and 
a broadening of the velocity distribution of the projectile. 
	The shift is expected to be a function of the impact 
parameter~\cite{Hufner84}. 
	The broadening is well represented by the Fermi momenta of the 
removed nucleons~\cite{Goldhaber74} and, within this picture, is
isotropic~\cite{Morrissey89}. 
	In the course of the deexcitation, the system is considered 
thermalised. 
	This justifies simplified assumptions on the angular 
distribution of the emitted particles. 
	After thermalisation, the angular distributions of the
particles emitted in sequential decay that reaches from evaporation 
of light particles up to symmetric fission, or produced by 
multifragmentation, are forward-backward symmetric in the system of 
the emitting source. 
	The angular anisotropy in the emission of light charged 
particles can be estimated within the Hauser-Feshbach 
picture~\cite{Hauser-Feshbach52}, the anisotropy of fission processes 
can be estimated in the framework of the statistical population of 
the K-quantum numbers~\cite{Vandenbosch-Huizenga73} and the 
anisotropic character of the multifragmentation process was already 
successfully described in early transport 
models~\cite{Stocker86,Aichelin91,Bonasera94}.
	However, since the angular momentum introduced in the collision 
phase is rather low and the excitation energy rather large, for the 
deconvolution procedure, we assume the angular distribution to be
approximately isotropic for any decay process, in the frame of the
decaying source. 
	In agreement with this expectation, we add that the reaction 
analysed in this work is in between a large range of incident 
energies for which signatures of isotropy were measured in several 
experiments: from lowly-excited fissioning systems~\cite{Sanders99,Moretto89} 
to spallation with high-energy light projectiles and relativistic 
heavy-ion collisions~\cite{Karnaukhov99,Karnaukhov03,Korteling90}).
	Still, we anticipate that we will find slight deviations from 
the isotropy in the data, which can not be described within our
assumptions.
	It may not be helpless to also remind 
the following fundamental characteristic of any inclusive 
experiment: an isotropic emission pattern does not correspond 
necessarily to a symmetric distribution of longitudinal velocity.
	The latter is determined by an asymmetry in the distribution of 
the source energy or impact parameters, while the isotropic 
emission is a characteristic related to one single source sorted 
out of the whole distribution of the sources.
	This is also the principle of our kinematic analysis which 
consists of a conventional deconvolution procedure allowing 
to extract longitudinal-velocity distributions which are not 
necessarily symmetric by using an assumption of isotropic emission.

\subsection{
    Deconvolution procedure
}
%
%
\begin{figure}[t!]\begin{center}
\includegraphics[angle=0, width=0.65\textwidth]{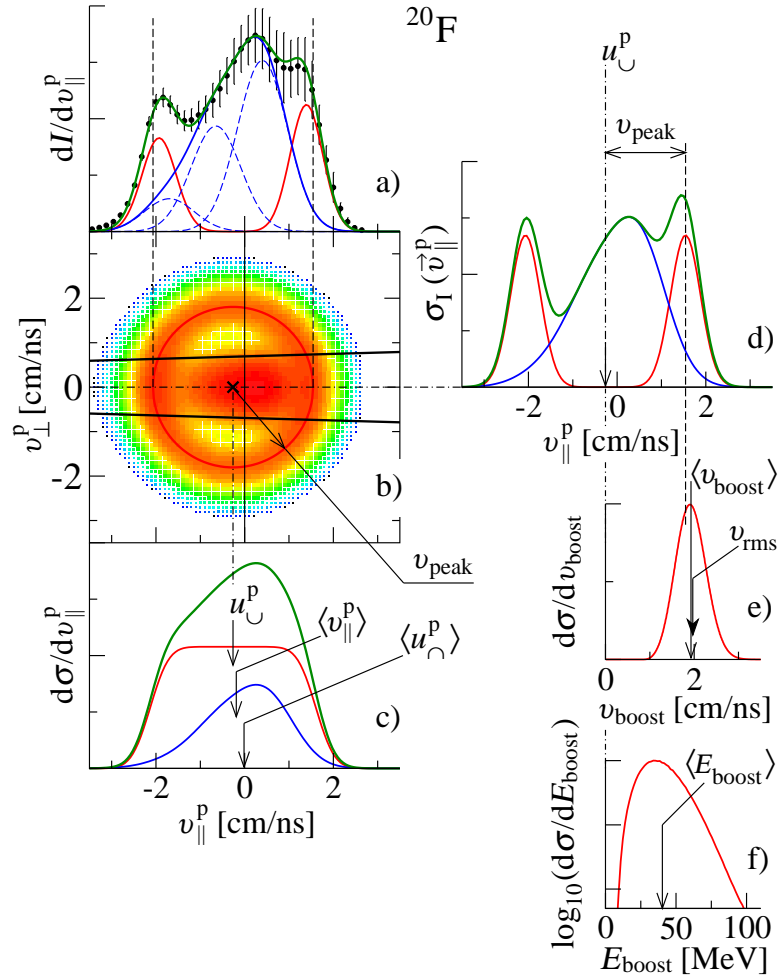}
\end{center}\caption
{
	Procedure of reconstruction of the invariant cross sections
	applied to the nuclide $^{20}$F. 
	The decomposition in the concave component of the Coulomb-hole mode
	and the convex component of the multifragmentation mode
	indicated by different line types as in fig.~\ref{fig2}.
(a)	Measured velocity distributions (black dots) and fitted
	spectra as in fig.~\ref{fig2}. The decomposition of
	the multifragmentation mode in three Gaussian components is shown.
(b)	Planar cut along the beam axis of the reconstructed full 
	distribution $\diff\sigma/\diff\vvec^\beam$ in the beam 
	frame as a cluster plot. The two lines indicate the 
	boundaries of the angular acceptance, inside of which the 
	fragments could be measured.
	The ridge of the Coulomb-hole component is indicated by the circle 
	of radius $v_{\mathrm{peak}}$.
(c)	Projection on the beam axis of the reconstructed full 
	distribution $\diff\sigma/\diff\vvec^\beam$. Arrows indicate 
	the positions of the source of the Coulomb-hole component
	$u_{\smallsmile}^\beam$, the average source position of the
	multifragmentation component $\langle u_{\smallfrown}^\beam\rangle$ 
	and the average longitudinal velocity $\langle\vpar^\beam\rangle$.
(d)	Invariant-cross-section distribution.
(e)	Reconstructed cross-section distribution for the 
	Coulomb-hole component as a function of the boost
	velocity in the source frame $v_{\mathrm{boost}}$.
	The plot allows to deduce the mean boost 
	$\langle v_{\mathrm{boost}}\rangle$, which differs from 
	$v_{\mathrm{peak}}$ and from $v_{\mathrm{rms}}$. 
(f)	Reconstructed cross-section distribution for the 
	Coulomb-hole component as a function of the total  
	kinetic energy in the source frame $E_{boost}$.
	$v_{\mathrm{rms}}$ is deduced from the mean value 
	$\langle E_{\mathrm{boost}}\rangle$.
}
\label{fig3}
\end{figure}
Once defined all this, we can go through the formalism, which is a
conventional deconvolution procedure.
	It is sketched for the nuclide
$^{20}$F in fig.~\ref{fig3} and is essentially based on 
the following three points, which resume the introductory discussion:
\begin{enumerate}
\item 
	Especially when dealing with relativistic protons, 
the kinematics of one individual nuclide can be 
decomposed in several isotropic emission patterns, each one centred 
around one apparent source $u$. 
\item 
	Each apparent source $u$ is localised at a longitudinal velocity
$u^\beam$ in the projectile frame (and $u^\Lab$ in the laboratory 
frame), which reflects the momentum transfer involved in the 
collision and is aligned along the beam axis: this assumption 
is equivalent to imposing that the overall kinematics is symmetric
with respect to the beam axis.
\item 
	The shape of the angular acceptance is known precisely.
\end{enumerate}
	The first two points are assumptions.
	The third point is an ingredient guaranteed by ion-optical 
calculations~\cite{Benlliure02}.
	As described in section~\ref{section2}, these
ion-optical calculations were also included in the procedure of 
raw-data treatment which had as an outcome the measured spectra of 
fig.~\ref{fig2}; they are validated by requiring that the 
different magnetic settings of the spectrometer do not produce
periodic structures on these spectra~\cite{Napolitani04}; 
such a requirement is a tool to check and even to refine the exact 
properties of the angular acceptance.

    We already defined the quantity $\diff I(\vpar^\beam)/\diff\vpar^\beam$
as the measured longitudinal-velocity spectrum in the projectile 
frame (see fig.~\ref{fig2}); it is a sum of the contributions of 
sources complying to the hypotheses (1) and (2), and is affected 
by the angular-acceptance shape of the spectrometer.
    Additionally, for a given number of sources of velocity 
$u_i^\beam$ in the projectile frame, since the velocities in the 
source frame are not relativistic, we define
\begin{equation}
    \sigma_{\mathrm{I}i}(\vpar^{\beam}-u^{\beam}_i) = 
        \frac{\diff\sigma}{\diff(\vvpar^{\beam}-\uvec^{\beam}_i)}
\label{eq1}\end{equation}
as the invariant cross section attributed to the source $i$.
    This is a production quantity which does not depend on the angular 
acceptance and, according to the assumption (1), it depends only
on the absolute value of the relative velocity $\vvpar^\beam-\uvec^\beam_i$
in the frame of the source $u^\beam_i$.

    These cross sections are obtained by the deconvolution of the
measured spectrum $\diff I(\vpar^\beam)/\diff\vpar^\beam$
taking into account the ion-optics required by the point (3).
    This procedure is detailed in a general form in appendix~\ref{appendixA}.
    By composing these contributions, we construct the 
invariant-velocity cross section accounting for all sources, 
independently of the angular acceptance, which is denoted by
\begin{equation}
    \si(\vvec^{\beam}) = \frac{\mathrm{c}^2}{\mathrm{m}^2}
    \frac{\diff\sigma}{\diff\vvec^{\beam}}\,
	.
\label{eq2}\end{equation}
As explained in appendix~\ref{appendixA}, we obtain the relation
\begin{equation}
    \si(\vvec^{\beam}) = \sum_i
    \sigma_{\mathrm{I}i}(|\vvec^{\beam}-\uvec^{\beam}_i|)\,
	.
\label{eq3}\end{equation}

	The spectrum reduces to a fully isotropic distribution
if all emitting sources coincide for one individual nuclide.

	This prescription was introduced in the simplified form of
isotropic emission from one unique source in the analysis of
the light residues in the system 
\FepGeV~\cite{Napolitani04}.
	For such system, no evidence was found for postulating a
spreading of emitting sources.
	On the contrary, the description of multiple-source
emission was necessary in the analysis of intermediate-mass 
fragments produced in the spallation of xenon.
%
\section{
    Results	\label{section4}
}
%
	 As shown in fig.~\ref{fig2}, the evolution of the measured
longitudinal-velocity spectra over a systematics of several
light nuclides ranging from lithium to silicon invites to 
disentangle two main components: 
one has a shape with a concave centre boarded by two maxima: it 
corresponds to a Coulomb hole; the other component has a convex 
shape with a tail extending in the backward direction: as discussed in the
following, we attribute to this mode the features of a multifragmentation
process.
	The decomposition of the measured longitudinal-velocity distribution 
in the Coulomb-hole mode (symbol $\smallsmile$) and the multifragmentation
mode (symbol $\smallfrown$) is shown in fig.~\ref{fig2} for elements 
ranging from lithium to silicon.
	The two kinematic modes are disentangleg by a fitting procedure which 
respects the constraints imposed by the detailed knowledge of the ion-optics.
	In the plot (a) of fig.~\ref{fig3} this decomposition is shown for $^{20}$F.
	The convex shape of the multifragmentation mode is fitted as a skewed 
Gaussian distribution and the concave shape of the Coulomb-hole mode is fitted 
as the superposition of two Gaussian shapes with integrals constrained to have 
a ratio compatible with the effect of the angular acceptance on the forward 
and backward emission kinematics.
	The forward side of the spectra of the heavier isotopes of 
beryllium were not fully measured or partially affected 
because the magnetic rigidity of the primary beam was circumvented in the experiment;
for these nuclides,
the constraint on the concave shape of the Coulomb-hole mode allowed
to profit from the precise measurement of the backward side, and the
parameters for the convex shape of the multifragmentation mode were deduced 
from isobaric nuclides.

	The double-humped shape of the Coulomb-hole component, with integral
$I_{\smallsmile} (\vpar^\beam)$, was interpreted as a shell in 
velocity space centred around one single isotropic emitting source
of recoil velocity $u_{\smallsmile}^\beam$ in the beam frame.
	The plot (b) of fig.~\ref{fig3} shows a planar cut along 
the beam direction of the reconstructed velocity distribution 
$\diff\sigma/\diff\vvec^\beam$; the radius $v_{\mathrm{peak}}$
indicated in the figure is the radius of the ridge of the ring marking 
the Coulomb-hole component.
	The same figure shows that the distribution 
$\diff\sigma/\diff\vvec^\beam$ is cut by the conical boundary of 
the acceptance so as to produce the two humps in the 
longitudinal-velocity spectra.
	The larger is the radius of the emission shell, the larger is 
the integral of the forward hump with respect to the backward hump.
	As the two peaks are spaced of a distance which largely exceeds 
their widths, their shapes are close to Gaussian distributions.

\subsection{Invariant cross sections} \label{section4A}
%
%
\begin{figure}[]\begin{center}
\includegraphics[angle=0, width=0.8\textwidth]{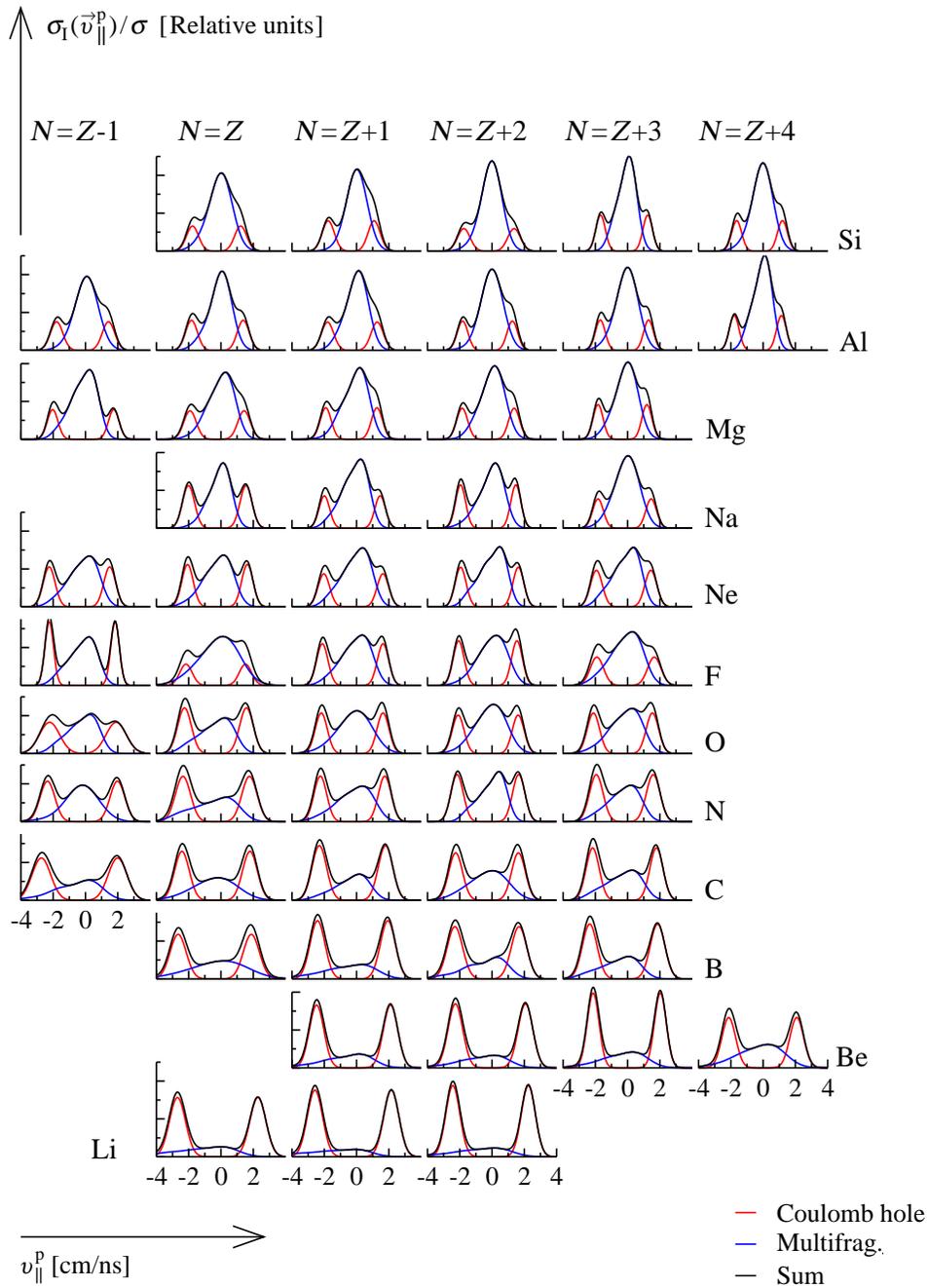}
\end{center}\caption
{
	Systematics of invariant-cross-section distributions
as a function of the nuclide for elements ranging from Li to Si
decomposed in the Coulomb-hole and multifragmentation components. 
}
\label{fig4}
\end{figure}
	We already anticipated that an isotropic emission pattern does 
not correspond necessarily to a symmetric distribution of 
longitudinal velocity.
	In other words, an asymmetric spectrum and an isotropic 
emission pattern are not two contradictory concepts.
	This is the case of the asymmetric shape of the 
multifragmentation component, of integral 
$I_{\smallfrown} (\vpar^\beam)$. 
	It could not be interpreted as the emission from one single 
isotropic source: we decomposed it in a set of Gaussian components 
$\diff I_{i\,\smallfrown} (\vpar^\beam) / {\diff\vpar^{\beam}}$
of different integrals, and different mean values
$u_{\,_{\scriptstyle i\,\smallfrown}}^\beam$.
	We assumed that the emission kinematics is independent of the 
recoil velocity of the corresponding source 
$u_{\,_{\scriptstyle i\,\smallfrown}}^\beam$: this assumption
results in assigning to all the Gaussian components the same width, 
which is determined by the forward side of the convex shape of the
multifragmentation mode (the effect of the angular acceptance on a 
Gaussian-like emission kinematics is discussed in ref.~\cite{Benlliure02}).
	In the plot (a) of fig.~\ref{fig3}, a decomposition of the
multifragmentation mode in three Gaussian components is shown; for this 
spectrum, a number of three Gaussians was chosen so that their 
spacing is smaller than their width, sufficiently to avoid the 
appearing of structures.

	By the fitting procedure, the measured spectrum is disentangled 
in the Coulomb-hole and multifragmentation components, so that
\begin{equation}
	\yield  =
		\frac{\diff I_{\smallsmile} (\vpar^\beam)}
			{\diff\vpar^{\beam}}
		+ \sum_i
		\frac{\diff I_{i\,\smallfrown} (\vpar^\beam)}
			{\diff\vpar^{\beam}}
	.
\label{eq4}\end{equation}
	All parameters involved in the fit have uncertainties deduced
from the corresponding covariance matrix.
	The plot (b) of fig.~\ref{fig3} confirms that the 
backward-forward asymmetry which characterises the Coulomb-hole 
component in the measured spectra is produced by the angular 
acceptance of the spectrometer and disappears in the reconstructed 
distribution. 
	On the contrary, the asymmetry of the multifragmentation component is not 
an effect of the acceptance and is kept in the reconstructed shape:
this confirms that, as we anticipated,
it depends on the reaction process.

	The projection of the reconstructed distribution 
$\diff\sigma/\diff\vvec^\beam$ on the beam axis is illustrated in 
the plot (c) of fig.~\ref{fig3}: it corresponds to the 
longitudinal-velocity distribution we would measure if the 
spectrometer had a full acceptance. In this representation the 
integrals of the different kinematic components give the 
corresponding production cross sections and their mean values 
give the average recoil of the corresponding sources.

	The variation of the reconstructed quantity 
$\diff\sigma/\diff\vvec^\beam$ along the beam axis coincides with
the distribution of the invariant cross section $\si(\vvpar^\beam)$
illustrated in the plot (d) of fig.~\ref{fig3}. 
	The Coulomb-hole component 
$\sigma_{\mathrm{I\,}_{\scriptstyle\smallsmile}}$ and the
multifragmentation components	
$\sigma_{\mathrm{I\,}_{\scriptstyle i\,\smallfrown}}$ of the 
distribution of invariant cross section are deduced from the
corresponding components of the measured longitudinal-velocity
spectra.
	The procedure of deconvolution requires the knowledge of 
the mean velocities of the sources 
$u_{\scriptstyle\smallsmile}^\beam$ and
$u_{\scriptstyle i\,\smallfrown}^\beam$.
	These quantities are however not measured directly, but deduced
through an optimisation procedure.
	The complete spectrum of invariant cross section is 
reconstructed by summing its components, shifted of the 
corresponding positions of the sources:
\begin{equation}
	\si(\vvpar^\beam)  =
		\sigma_{\mathrm{I\,}_{\scriptstyle\smallsmile}}
		(\vpar^{\beam}-u_{\scriptstyle\smallsmile}^\beam)\,
		+ \sum_i
		\sigma_{\mathrm{I\,}_{\scriptstyle i\,\smallfrown}}
		(\vpar^{\beam}-u_{\scriptstyle\smallfrown}^\beam)\,
	,
\label{eq5}\end{equation}
    This is the application of Eq.~(\ref{eq2}) when $\vvec^{\beam}$
is constrained to vary along the beam axis.
%
	The procedure of velocity reconstruction was applied for all 
intermediate-mass fragments ranging from lithium to silicon.
	In fig.~\ref{fig4} the resulting systematics of 
invariant-cross-section distributions is shown, deduced from the
deconvolution of the measured longitudinal-velocity spectra
illustrated in fig.~\ref{fig2}.
	The main parameters characterising the reconstructed 
spectra are listed in table~\ref{tab3} in the appendix~\ref{appendixB}.
%
%
%

	One more representation of the Coulomb-hole component is obtained by 
integrating the corresponding reconstructed distribution
$\diff\sigma/\diff\vvec^\beam$ over the polar angle.
	This is shown in the plot (e) of fig.~\ref{fig3}; it
probes the evolution of the cross section as a function of the 
Coulomb boost $v_{\mathrm{boost}}$. 
	The plot (f) of fig.~\ref{fig3} is an equivalent 
representation in terms of the total kinetic energy $E_{boost}$.
	The mean value $\langle E_{\mathrm{boost}}\rangle$ is indicated
in the plot (f) and corresponds to the velocity $v_{\mathrm{rms}}$
indicated in the plot (e). 
	$v_{\mathrm{rms}} = \sqrt{\langle v_{\mathrm{boost}}\rangle^2+V}$ 
accounts for the variance $V$ of the velocity distribution and is slightly
larger than the mean velocity $\langle v_{\mathrm{boost}}\rangle$, also
indicated in the plot (e). 
	In section \ref{section5A} the quantity $v_{\mathrm{rms}}$ is compared
to fission velocities calculated under different prescriptions for the 
Coulomb boost.
\subsection{Production cross sections} \label{section4B}
%
%
\begin{figure}[b]\begin{center}
\includegraphics[angle=0, width=0.6\textwidth]{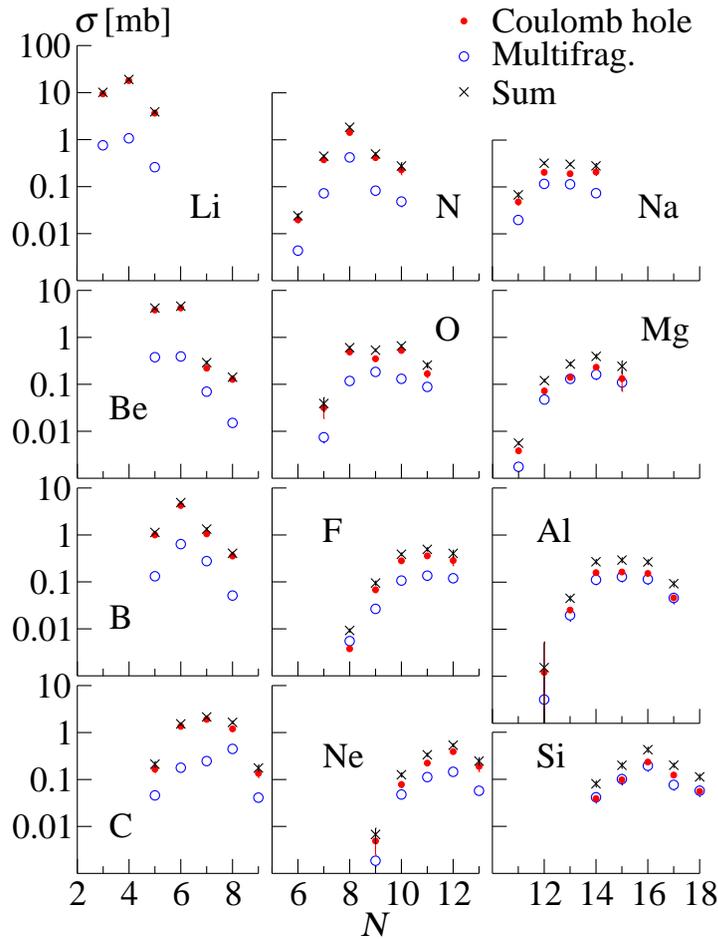}
\end{center}\caption
{
	Production cross sections for the isotopes of elements
	ranging from Li to Si.
	The contributions from the two kinematic modes are shown.
}
\label{fig5}
\end{figure}
%
%
\begin{figure}[]\begin{center}
\includegraphics[angle=0, width=0.55\textwidth]{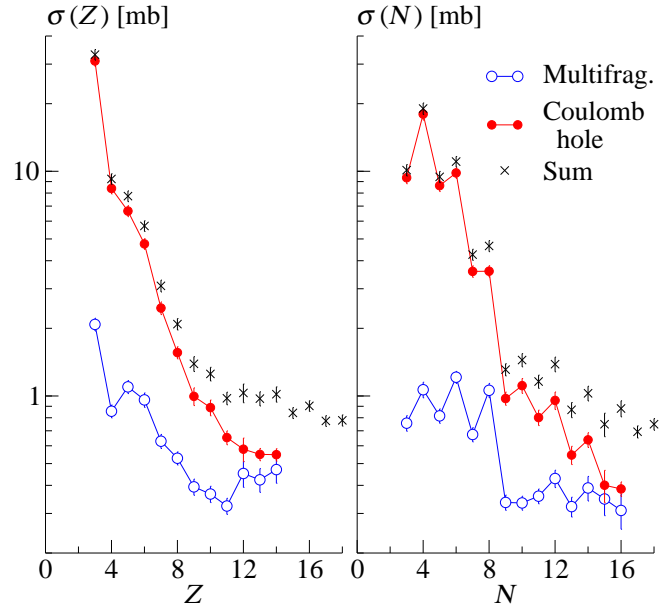}
\end{center}\caption
{
	Element (left) and neutron-number (right) distributions
of the intermediate-mass-fragment cross sections.
	The contribution from the two kinematic modes are indicated.
}
\label{fig6}
\end{figure}
%
%
\begin{figure}[]\begin{center}
\includegraphics[angle=0, width=0.55\textwidth]{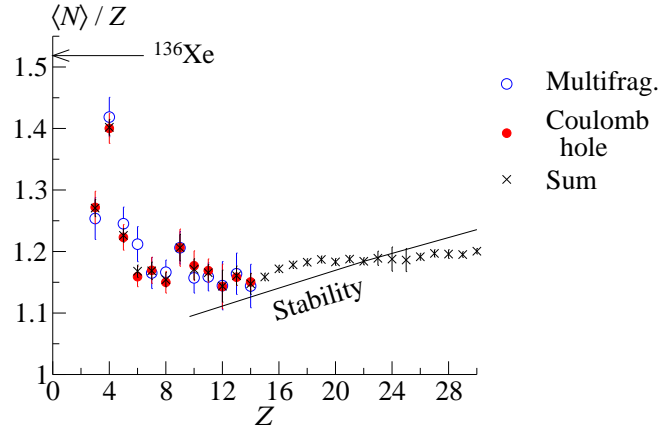}
\end{center}\caption
{
	Evolution of the average neutron-to-proton ratio 
$\langle N/Z\rangle$ of the intermediate-mass fragments as a 
function of the element number.
	The contributions from the two kinematic modes are indicated.
}
\label{fig7}
\end{figure}
	The production cross sections were deduced from integrating the
reconstructed distributions $\diff\sigma/\diff\vvec^\beam$.
	The disentangling of the velocity spectra into the two 
kinematic modes allowed to separate the contributions of each mode.
	The results are presented for the isotopic chains of elements 
ranging from lithium to silicon in fig.~\ref{fig5} and listed in
table~\ref{tab1}.
	The uncertainties account for statistical and systematical 
contributions (see ref.~\cite{Napolitani07} for details).
	The production of the lightest elements is mostly related to the 
Coulomb-hole mode. 
	For heavier elements the multifragmentation mode gains gradually a larger
fraction of cross section and it equals the Coulomb-hole mode around
silicon, where the method of deconvolution reaches its limit of
applicability.
%
%
\begin{table}[th!]
\caption
{
\label{tab1}
	Nuclide cross sections $\sigma$ measured in this work for 
the production of elements ranging from Li to Si.
The cross section restricted to the Coulomb-hole or multifragmentation  
mode is indicated as $\sigma_{\smallsmile}$ or 
$\sigma_{\smallfrown}$, respectively.
}
\vspace{5pt}
\begin{tabular}{l l}
$\quad A \quad \sigma $ [mb] $ \qquad \sigma_{\smallsmile} $ [mb] $ \qquad \sigma_{\smallfrown} $ [mb] &
$\quad A \quad \sigma $ [mb] $ \qquad \sigma_{\smallsmile} $ [mb] $ \qquad \sigma_{\smallfrown} $ [mb]
\vspace{3pt}\\
\hline\vspace{-3pt}\\
\begin{tabular}{l l l}
 $\qquad$                                    Li  \\
$\,\,\,   6\;\;     10.12(59)  $ & $      9.36(56)  $ & $     0.757(60)  $\\
$\,\,\,   7\;\;     19.0\pm1.1  $ & $     18.0\pm1.1  $ & $  1.066(85)  $\\
$\,\,\,   8\;\;      3.90(30)  $ & $      3.64(29)  $ & $     0.259(33)  $\\
 $\qquad$                                    Be  \\
$\,\,\,   9\;\;      4.20(25)  $ & $      3.82(23)  $ & $     0.378(32)  $\\
$\,\,\,  10\;\;      4.61(28)  $ & $      4.22(26)  $ & $     0.392(32)  $\\
$\,\,\,  11\;\;      0.290(23)  $ & $     0.220(19)  $ & $     0.070(10)  $\\
$\,\,\,  12\;\;      0.142(10)  $ & $     0.1267(94)  $ & $    0.0151(20)  $\\
 $\qquad$                                    B   \\
$\,\,\,  10\;\;      1.131(73)  $ & $     0.999(66)  $ & $     0.132(18)  $\\
$\,\,\,  11\;\;      4.88(29)  $ & $      4.25(26)  $ & $     0.638(46)  $\\
$\,\,\,  12\;\;      1.33(10)  $ & $      1.055(89)  $ & $     0.277(32)  $\\
$\,\,\,  13\;\;      0.406(27)  $ & $     0.355(25)  $ & $    0.0512(55)  $\\
 $\qquad$                                    C   \\
$\,\,\,  11\;\;      0.212(28)  $ & $     0.166(26)  $ & $    0.0459(75)  $\\
$\,\,\,  12\;\;      1.528(92)  $ & $      1.350(84)  $ & $     0.178(13)  $\\
$\,\,\,  13\;\;      2.15(13)  $ & $      1.90(12)  $ & $     0.247(27)  $\\
$\,\,\,  14\;\;      1.65(12)  $ & $      1.197(89)  $ & $     0.449(52)  $\\
$\,\,\,  15\;\;      0.176(27)  $ & $     0.135(25)  $ & $    0.0412(80)  $\\
 $\qquad$                                    N   \\
$\,\,\,  13\;\;      0.0241(28)  $ & $    0.0197(27)  $ & $   0.00436(59)  $\\
$\,\,\,  14\;\;      0.446(33)  $ & $     0.374(30)  $ & $    0.0723(62)  $\\
$\,\,\,  15\;\;      1.85(13)  $ & $      1.42(11)  $ & $     0.422(36)  $\\
$\,\,\,  16\;\;      0.499(38)  $ & $     0.416(35)  $ & $     0.0823(87)  $\\
$\,\,\,  17\;\;      0.276(50)  $ & $     0.228(48)  $ & $     0.048(11)  $\\
 $\qquad$                                    O   \\
$\,\,\,  15\;\;      0.039(13)  $ & $     0.032(13)  $ & $    0.0075(18)  $\\
$\,\,\,  16\;\;      0.603(40)  $ & $     0.485(35)  $ & $     0.118(10)  $\\
$\,\,\,  17\;\;      0.534(38)  $ & $     0.350(28)  $ & $     0.184(19)  $\\
$\,\,\,  18\;\;      0.656(50)  $ & $     0.525(45)  $ & $     0.131(12)  $\\
$\,\,\,  19\;\;      0.256(33)  $ & $     0.168(30)  $ & $     0.088(11)  $\\
 $\qquad$                                    F   \\
$\,\,\,  17\;\;      0.00925(79)  $ & $   0.00376(23)  $ & $   0.00549(69)  $\\
$\,\,\,  18\;\;      0.095(10)  $ & $     0.0678(91)  $ & $    0.0266(28)  $\\
\end{tabular}&\begin{tabular}{l l l}
$\,\,\,  19\;\;      0.388(28)  $ & $     0.281(23)  $ & $     0.1066(93)  $\\
$\,\,\,  20\;\;      0.494(44)  $ & $     0.358(39)  $ & $     0.136(13)  $\\
$\,\,\,  21\;\;      0.404(70)  $ & $     0.285(65)  $ & $     0.119(22)  $\\
 $\qquad$                                    Ne  \\
$\,\,\,  19\;\;      0.0068(24)  $ & $    0.0049(24)  $ & $   0.00186(47)  $\\
$\,\,\,  20\;\;      0.126(13)  $ & $     0.078(12)  $ & $    0.0479(46)  $\\
$\,\,\,  21\;\;      0.336(28)  $ & $     0.223(21)  $ & $     0.113(14)  $\\
$\,\,\,  22\;\;      0.538(44)  $ & $     0.392(36)  $ & $     0.146(19)  $\\
$\,\,\,  23\;\;      0.247(45)  $ & $     0.189(43)  $ & $     0.058(11)  $\\
 $\qquad$                                    Na  \\
$\,\,\,  22\;\;      0.0677(80)  $ & $    0.0480(71)  $ & $    0.0197(28)  $\\
$\,\,\,  23\;\;      0.321(22)  $ & $     0.205(13)  $ & $     0.116(14)  $\\
$\,\,\,  24\;\;      0.305(21)  $ & $     0.191(12)  $ & $     0.114(14)  $\\
$\,\,\,  25\;\;      0.282(35)  $ & $     0.208(32)  $ & $     0.074(10)  $\\
 $\qquad$                                    Mg  \\
$\,\,\,  23\;\;      0.00559(46)  $ & $   0.00384(29)  $ & $   0.00175(29)  $\\
$\,\,\,  24\;\;      0.1196(91)  $ & $    0.0723(46)  $ & $    0.0473(65)  $\\
$\,\,\,  25\;\;      0.271(27)  $ & $     0.1400(87)  $ & $     0.131(23)  $\\
$\,\,\,  26\;\;      0.394(44)  $ & $     0.232(20)  $ & $     0.162(36)  $\\
$\,\,\,  27\;\;      0.242(72)  $ & $     0.132(62)  $ & $     0.110(36)  $\\
 $\qquad$                                    Al  \\
$\,\,\,  25\;\;      0.0015(38)  $ & $    0.0012(38)  $ & $   0.00032(63)  $\\
$\,\,\,  26\;\;      0.0453(56)  $ & $    0.0255(22)  $ & $    0.0197(49)  $\\
$\,\,\,  27\;\;      0.271(29)  $ & $     0.159(14)  $ & $     0.112(24)  $\\
$\,\,\,  28\;\;      0.294(34)  $ & $     0.164(14)  $ & $     0.130(29)  $\\
$\,\,\,  29\;\;      0.268(31)  $ & $     0.153(14)  $ & $     0.115(26)  $\\
$\,\,\,  30\;\;      0.092(13)  $ & $    0.0463(38)  $ & $     0.046(12)  $\\
 $\qquad$                                    Si  \\
$\,\,\,  28\;\;      0.080(11)  $ & $    0.0388(33)  $ & $     0.042(10)  $\\
$\,\,\,  29\;\;      0.199(27)  $ & $    0.0977(81)  $ & $     0.101(25)  $\\
$\,\,\,  30\;\;      0.427(53)  $ & $     0.233(21)  $ & $     0.194(46)  $\\
$\,\,\,  31\;\;      0.199(22)  $ & $     0.123(10)  $ & $     0.076(18)  $\\
$\,\,\,  32\;\;      0.113(16)  $ & $    0.0555(47)  $ & $     0.057(14)  $\\
$\, $ & $\, $ & $\, $\\
\end{tabular}
\\
\end{tabular}
\end{table}
	This evolution can be appreciated in the 
element and neutron-number distributions
of cross sections in fig.~\ref{fig6}. 
	The average neutron-to-proton ratio $\langle N/Z\rangle$
of the intermediate-mass fragments produced by the two modes are 
indistinguishable, as shown in fig.~\ref{fig7}: both the two 
modes feed prevalently the neutron-rich side of the nuclide chart 
with very similar shapes of the cross-section distributions.
\subsection{Limitations of the method} \label{section4C}
	It was not possible to extend the deconvolution for
heavier elements than silicon because the uncertainties of the
fitting parameters for the identification of the two kinematic
modes in the measured velocity spectra determined a limit for the
analysis.
	Above silicon, a simplified method was employed, based on 
the assumption of one unique isotropic source associated to the 
kinematics of one fragment~\cite{Napolitani07}; the same 
simplification was also employed in the analysis of the kinematics
of the intermediate-mass fragments formed in the system 
\FepGeV~\cite{Napolitani04}.

%
	Concerning the multifragmentation mode, we already pointed
out the presence of a tail extending to the backward direction
which characterises the lightest fragments.
	Such tail should result from the early stages of the
collision and reflects fluctuations in the momentum transfer:
following an intranuclear-cascade picture, these fluctuations are
connected to a smaller or larger interaction of the shower of
ejectile nucleons with the system, and result in a distribution of 
emitting sources.
	This distribution would be attributed to the longitudinal 
momentum transfer due to friction, while the shape of the isotropic
distribution would be governed by the Fermi momenta of the emitted 
nucleons~\cite{Goldhaber74} and the recoil in the decay process 
(fission or multifragmentation).
	It must be admitted, however, that the relatively small 
angular acceptance does not allow to get direct experimental 
information on the shape of the velocity distribution in transverse 
direction.
	Guided by the present knowledge from other experiments and 
in agreement with the theoretical expectations for the 
contributions from Fermi momenta and decay recoil, we still keep
the underlying assumption of isotropic emission patterns as a 
technical option to describe the tail to backward velocities.
	However, the validity of this assumption may only be tested 
by a full-acceptance experiment.

%
\section{
    Discussion
\label{section5}
}
%
	In the analysis of the velocity spectra of the 
intermediate-mass fragments formed in the system 
\FepGeV~\cite{Napolitani04} similar 
results were found: the effect of the Coulomb field on the 
kinematics of fragments reveals the interplay of a Coulomb-hole mode and a
multifragmentation mode, as shown in the left column of fig.~\ref{fig8}.
	However, in such system the Coulomb-hole mode exhibits two  
humps with large widths, with the consequence that the two modes could 
not be disentangled in single spectra and they were identified only 
for the extreme case when the contribution of one component is dominant; 
their coexistence in the production of one individual nuclide could only be 
postulated on the basis of a model calculation, as shown in the right column 
of fig.~\ref{fig8}.
	The calculation, where the distribution of hot fragments is
evaluated with the intranuclear exciton-cascade 
model~\cite{Gudima83}, and their breakup is simulated by the
Statistical Multifragmentation Model~\cite{Botvina85b,Botvina90,Bondorf95},
associates the Coulomb-hole component in the velocity spectrum
to events with multiplicity of larger fragments than alpha $M=2$ and the 
multifragmentation component to events with multiplicity $M>2$.
%
%
%
\begin{figure}[b]\begin{center}
\includegraphics[angle=0, width=0.55\textwidth]{Fig08.eps}
\end{center}\caption
{
	Left column. Measured velocity distributions and invariant-cross-section 
distributions for four nuclides produced in
\FepGeV.
	Right column. Invariant-cross-section distribution for $^{6}$Li,
calculated with the intranuclear exciton-cascade 
model~\cite{Gudima83}, coupled with the Statistical Multifragmentation 
Model~\cite{Botvina85b,Botvina90,Bondorf95}; two components, associated to
the multiplicity $M=2$ and $M>2$ (for larger fragments than alpha) 
are indicated.
	Data and calculation from~\cite{Napolitani04}.
}
\label{fig8}
\end{figure}
%
%
%

	On the contrary, for the heavier system 
$^{136}$Xe$_{(1\,A\,\textrm{GeV})}+p$, the identification of the
two distinct channels in the formation of 
intermediate-mass fragments is an experimental result, even when 
they coexist in the formation of the same nuclide.
	To our knowledge, there have not been previous attempts to
extract this observable from inclusive experiments and for 
similar systems.
\subsection{Exit channel} \label{section5A}
\subsubsection{Coulomb hole.}
\begin{figure}[b]\begin{center}
\includegraphics[angle=0, width=0.6\textwidth]{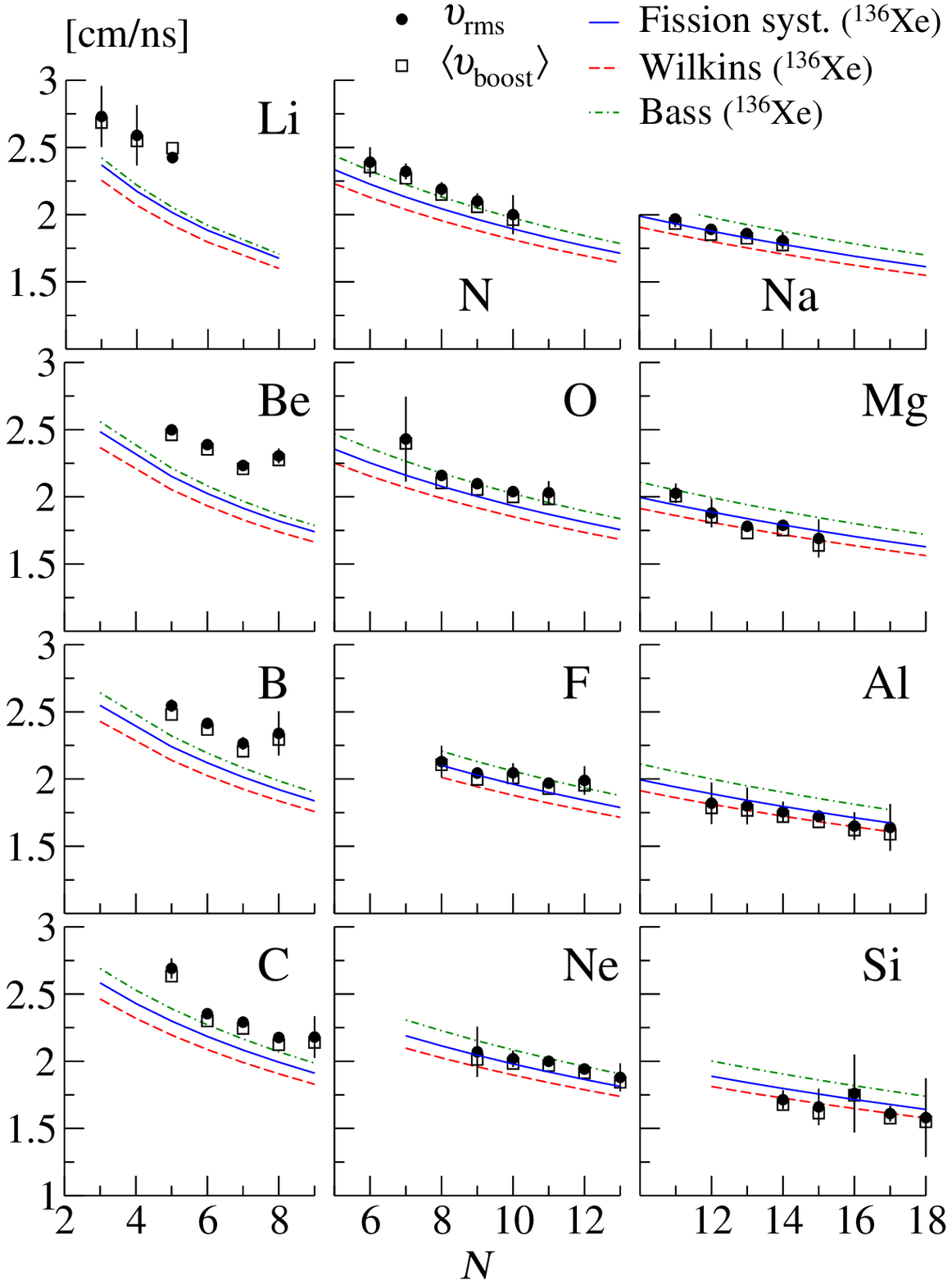}
\end{center}\caption
{
	Evolution of the reconstructed boost $v_{\mathrm{rms}}$ of the Coulomb-hole 
mode (and of the mean boost $\langle v_{\mathrm{boost}}\rangle$, 
plotted for comparison, without error bars)
as a function of the neutron number, for elements ranging from 
lithium to silicon.
	Three prescriptions for calculating the Coulomb boost in the 
binary split of $^{136}$Xe are compared to the measurement: 
the total-kinetic-energy systematics of 
Beck and Szanto de Toledo~\cite{Beck96},
the scission-point model of Wilkins et 
al.~\cite{Wilkins76, Bockstiegel97} 
and the nucleus-nucleus fusion
model of Bass~\cite{Bass79, Bass80}.
}
\label{fig9}
\end{figure}

	Double-humped spectra were measured at the FRagment Separator
in several experiments: they were usually taken as the signal of 
binary fission characterised by mostly symmetric splits (without 
being exhaustive, we cite refs.~\cite{Enqvist01, Bernas03}), and 
were also observed in presence of asymmetric fission splits, when light 
fragments were measured~\cite{Ricciardi06}.
    With the purpose of comparing to fission data 
it is therefore interesting to quantify the Coulomb repulsion 
reflected by the forward and backward peaks in the present data.
	In fig.~\ref{fig9}, the boost $v_{\mathrm{rms}}$
is deduced as in the plot (e) of fig.~\ref{fig3} and its evolution 
is studied as a function of the neutron number for elements ranging 
from lithium to silicon.
	The same experimental points are collected in table~\ref{tab2}.
	As described in section \ref{section4A}, $v_{\mathrm{rms}}$ is
the velocity deduced from the mean value of the total kinetic energy 
distribution $\langle E_{\mathrm{boost}}\rangle$, which represents the
average mechanism leading to the observed nucleus.
	Because of the wide velocity spectra, this quantity results slightly 
larger than the mean value of the velocity distribution 
$\langle v_{\mathrm{boost}}\rangle$, which is also plotted
in fig.~\ref{fig9} for comparison.
	In the figure, the reconstructed boost $v_{\mathrm{rms}}$ is compared to 
the fission velocity calculated under different prescriptions, all
applied to the same extreme case:
the binary split of the largest possible mother nucleus
$^{136}$Xe, which also determines the largest boost. 
    The calculations do not change sensibly for the split of lighter 
isotopes of xenon, if fission occurred after the emission of some neutrons.
	Three prescriptions are tested.
	First, we compare to the empirical systematics of total kinetic energy 
for light fissioning nuclei condensed in the parametrisation of 
Beck and Szanto de Toledo~\cite{Beck96}, and further adapted to describe asymmetric splits
by using the prescription of Ref.~\cite{Napolitani04}.
%
%
%
%
	Then, we compare to the scission-point model of Wilkins et al., 
which was already used to describe similar data~\cite{Ricciardi06}.
	The model deduces the total kinetic energy from an empirical 
liquid-drop description, based on the parametrisation of 
Refs.~\cite{Wilkins76, Bockstiegel97}.
	The underlying picture of the model, which is the fission of
a deformed system through the formation of a neck, is
appropriate to describe the symmetric split of heavy fissioning 
systems, but is not sufficiently justified when the formation of
intermediate-mass fragments is described.
	This second calculation may be regarded as a lower limit.
	As a third option, we deduce the boost from the nuclear potential which
the light fragment and its heavy partner would explore in a fusion
reaction, according to the empirical model of 
Bass~\cite{Bass79, Bass80}: within this more appropriate 
description, the Coulomb repulsion acts on non-deformed fission 
fragments and results in the largest boost, compared to the 
first two prescriptions.
	For the extreme case of a fissioning $^{136}$Xe, while isotopes 
from nitrogen to neon are well described both by the fission 
systematics and the Bass potential, the heavier isotopes from sodium 
to silicon could be compatible with a lighter fissioning system, 
according to the same two prescriptions, or still with the split of  
$^{136}$Xe within the scission-point model of Wilkins et al.
	The large increase of the boost $v_{\mathrm{rms}}$ for decreasing element 
numbers in the region from lithium to carbon, is not even described 
by the Bass potential.
	This phenomenology is rather general:
the same conclusion was presented for the analysis of the 
system $^{56}$Fe$_{(1\,A\,\textrm{GeV})}+p$~\cite{Napolitani04} and
a similar tendency was found for the fission fragments in
the system $^{238}$U$_{(1\,A\,\textrm{GeV})}+p$~\cite{Ricciardi06}.
%
%
\begin{table}[t]
\caption
{
\label{tab2}
	Reconstructed boost $v_{\mathrm{rms}}$, deduced for the Coulomb-hole mode. 
	The uncertainty, indicated in parenthesis, includes both 
statistical and systematical errors.
}
\begin{tabular}{l l l l}
$\;\;\, A $
$\;\; v_{\mathrm{rms}}$
&
$\;\;\, A $
$\;\; v_{\mathrm{rms}}$
&
$\;\;\, A $
$\;\; v_{\mathrm{rms}}$
&
$\;\;\, A $
$\;\; v_{\mathrm{rms}}$
\vspace{3pt}\\
\hline\vspace{-3pt}\\
\begin{tabular}{l l}
$\qquad$   Li \\
$\;\;6\;\;  2.73(23)  $\\
$\;\;7\;\;  2.59(22)  $\\
$\;\;8\;\;  2.424(43) $\\
$\qquad$   Be \\
$\;\;9\;\;  2.498(15) $\\
$ 10\;\;  2.388(20) $\\
$ 11\;\;  2.233(42) $\\
$ 12\;\;  2.305(55) $\\
$\qquad$   B \\
$ 10\;\;  2.546(47) $\\
$ 11\;\;  2.414(20) $\\
$ 12\;\;  2.265(48) $\\
$ 13\;\;  2.34(16)  $\\
$\qquad$   C \\
$ 11\;\;  2.691(75) $\\
$ 12\;\;  2.354(23) $\\
\end{tabular}&\begin{tabular}{l l}
$\,\, 13\;\;  2.291(31) $\\
$\,\, 14\;\;  2.176(37) $\\
$\,\, 15\;\;  2.18(16)  $\\
$\qquad$   N \\
$\,\, 13\;\;  2.39(11)  $\\
$\,\, 14\;\;  2.321(58) $\\
$\,\, 15\;\;  2.190(51) $\\
$\,\, 16\;\;  2.101(57) $\\
$\,\, 17\;\;  2.00(15)  $\\
$\qquad$   O \\
$\,\, 15\;\;  2.43(32)  $\\
$\,\, 16\;\;  2.159(35) $\\
$\,\, 17\;\;  2.098(21) $\\
$\,\, 18\;\;  2.039(33) $\\
$\,\, 19\;\;  2.031(86) $\\
$\qquad$   F \\
$\,\, 17\;\;  2.13(12)  $\\
\end{tabular}&\begin{tabular}{l l}
$\,\, 18\;\;  2.045(44) $\\
$\,\, 19\;\;  2.046(70) $\\
$\,\, 20\;\;  1.969(39) $\\
$\,\, 21\;\;  1.99(11)  $\\
$\qquad$   Ne \\
$\,\, 19\;\;  2.07(19)  $\\
$\,\, 20\;\;  2.019(60) $\\
$\,\, 21\;\;  2.000(32) $\\
$\,\, 22\;\;  1.942(32) $\\
$\,\, 23\;\;  1.88(11)  $\\
$\qquad$   Na \\
$\,\, 22\;\;  1.969(59) $\\
$\,\, 23\;\;  1.892(16) $\\
$\,\, 24\;\;  1.860(20) $\\
$\,\, 25\;\;  1.808(62) $\\
$\qquad$   Mg \\
$\,\, 23\;\;  2.026(73) $\\
\end{tabular}&\begin{tabular}{l l}
$\,\, 24\;\;  1.88(11)  $\\
$\,\, 25\;\;  1.781(37) $\\
$\,\, 26\;\;  1.788(40) $\\
$\,\, 27\;\;  1.69(14)  $\\
$\qquad$   Al \\
$\,\, 25\;\;  1.82(15)  $\\
$\,\, 26\;\;  1.80(14)  $\\
$\,\, 27\;\;  1.756(76) $\\
$\,\, 28\;\;  1.723(45) $\\
$\,\, 29\;\;  1.65(10)  $\\
$\,\, 30\;\;  1.64(17)  $\\
$\qquad$   Si \\
$\,\, 28\;\;  1.714(71) $\\
$\,\, 29\;\;  1.66(14)  $\\
$\,\, 30\;\;  1.76(29)  $\\
$\,\, 31\;\;  1.610(59) $\\
$\,\, 32\;\;  1.58(29)  $\\
\end{tabular}
\\
\end{tabular}
\end{table}
	Even the simultaneous emission of a third light fragment or 
particle would not deform the shape of the Coulomb-hole component 
sensibly.
	The large value of the Coulomb boost $v_{\mathrm{rms}}$ may 
manifest the presence of some additional contribution to the total 
kinetic energy.

	As a general conclusion of this section, we relate the Coulomb-hole mode 
to events with low fragment multiplicity and large mass asymmetry 
$(A_1 - A_2)(A_1 + A_2)$, where $A_1$ is the mass number of the 
heaviest fragment and $A_2$ indicates the second heaviest fragment,
the one the velocity spectrum belongs to.
	The large width of the backward and forward peaks 
$\langle\sigma_{\| \textrm{peak}}^\beam\rangle$, plotted 
in fig.~\ref{fig10} for $A\le 30$, is compatible with a large 
number of decay paths all ending in the production of the observed 
nuclide; these decay paths account for several different breakup 
configurations and different quantities of particles which, 
depending on the excitation energy of the system, can be evaporated 
before and after the breakup.

\subsubsection{Multifragmentation}
	The multifragmentation mode extends over a large range of recoil 
velocities and indicates that the kinetic energy is shared among a large 
multiplicity of fragments with small mass asymmetry.
	In the analysis of the system  
$^{56}$Fe$_{(1\,A\,\textrm{GeV})}+p$ such a shape was proposed as
a signature compatible with multifragmentation.
	In such system, which has a larger excitation energy per nucleon 
than $^{136}$Xe$+p$, the multifragmentation mode was found to predominate 
already above carbon~\cite{Napolitani04}.
	Along the same line, in agreement with the picture that the 
multifragmentation share increases with the excitation of the 
system, inclusive measurements at the incident energy of 1 GeV per 
nucleon found only a small contribution from the Coulomb-hole component 
in comparison with the multifragmentation mode: it was the case of the systems 
$^{56}$Fe$+^{\textrm{nat}}$Ti~\cite{Napolitani04},
$^{136}$Xe$+^{\textrm{nat}}$Ti~\cite{Napolitani04b},
$^{136}$Xe$+^{\textrm{208}}$Pb~\cite{Henzlova}, and
$^{124}$Xe$+^{\textrm{208}}$Pb~\cite{Henzlova}).
%

%
\subsubsection{Widths of the velocity spectra}
%
%
%
\begin{figure}[t]\begin{center}
\includegraphics[angle=0, width=0.6\textwidth]{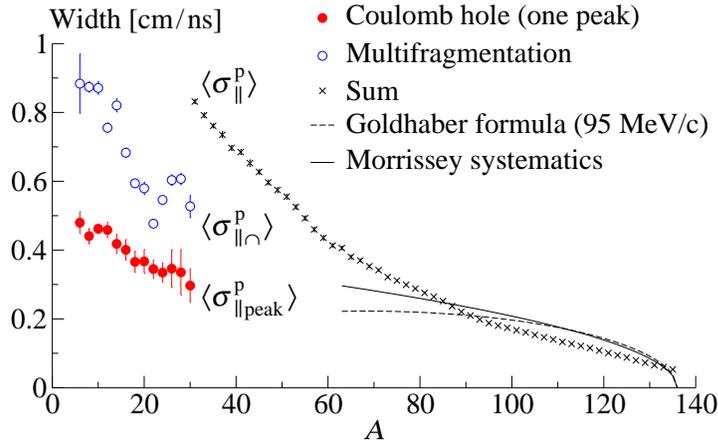}
\end{center}\caption
{
	Evolution as a function of the mass number of the residue of 
the mean standard deviation of the invariant-cross-section distribution averaged on 
the atomic number. 
	For lower masses than $A=30$ the width associated to the 
forward side of the multifragmentation component and the width of each peak of 
the Coulomb-hole mode are shown.
	The mean standard deviation of the full spectrum is shown for masses heavier than 
$A=30$.
	The systematics of Morrissey\cite{Morrissey89} and the 
prescription of Goldhaber~\cite{Goldhaber74} are shown for 
comparison.
}
\label{fig10}
\end{figure}
	In the system $^{136}$Xe$_{(1\,A\,\textrm{GeV})}+p$, the
contribution of the multifragmentation mode prevails gradually on the Coulomb-hole 
mode for elements of larger mass number till, around silicon, the unfolding of 
the two components becomes too arduous.
	For silicon nuclides, the contribution of the Coulomb-hole mode is still about
half the production cross section of silicon and should still 
survive for a long range of elements.
	In this respect, the presence of the Coulomb-hole mode is still appreciable in the 
widths of the longitudinal-velocity spectra.
	In fig.~\ref{fig10}, the evolution of the mean standard deviation 
$\langle\sigma_{\|}^\beam\rangle$ of the 
invariant-cross-section distribution is shown as a function of the mass 
number of the residue for $A>30$.
	For $A\le 30$ the mean standard deviation extracted from the forward side of
the multifragmentation mode in the invariant-cross-section representation is 
plotted: this standard deviation is equal to the average standard deviation
$\langle\sigma_{\|\,_{\scriptstyle i\,\smallfrown}}^\beam\rangle$
of the components of the multifragmentation mode.
	Around silicon, the shape of the multifragmentation mode loses its tail and
the width extracted from the forward side coincides with the width of the full 
multifragmentation mode.
	The mismatch at $A=30$ between the widths of the multifragmentation 
component and of the full velocity distribution confirms that the 
contribution of the Coulomb-hole component is still large.

%
%
\begin{figure}[t]\begin{center}
\includegraphics[angle=0, width=0.6\textwidth]{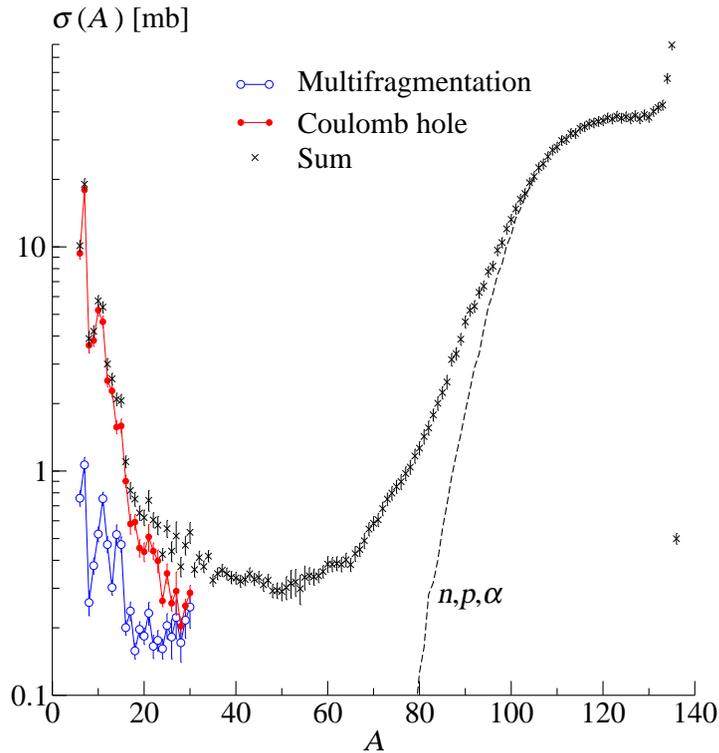}
\end{center}\caption
{
	Mass-number distributions of the residue-production cross.
	For the light fragments the contribution from the two
kinematic modes are indicated.
	A Weisskopf calculation for the 
expected contribution of the residues which evaporated neutrons, 
protons and alpha particles is shown. 
}
\label{fig11}
\end{figure}
\subsubsection{Nuclide production}
	A survey on the production cross section, taken from 
ref.~\cite{Napolitani07}, is presented in fig.~\ref{fig11}.
	The contributions of the Coulomb-hole and multifragmentation modes, 
deduced in this work, are indicated; a calculation is also included in the 
figure to estimate the contribution of a Weisskopf 
evaporation chain restricted to the emission of only neutrons, 
protons and alpha particles.
	Combining the production and the 
kinematic observables, we can estimate that in the region of
mass numbers corresponding to the minimum of cross section, while 
the information on the Coulomb-hole component gradually vanishes and the 
production cross section of intermediate-mass fragments drops 
steeply, the yield of the evaporation residues rises and the 
kinematics reflected in the velocity spectra is replaced by the 
characteristic Gaussian shape of the evaporation residues.
	Above half the mass number of the projectile we observe only 
the kinematics of the heaviest fragments.
	When approaching the mass of the projectile, as shown in 
fig.~\ref{fig10}, the velocity width of the invariant cross-section 
distributions evolves in qualitative agreement with
the empirical systematics of Morrissey\cite{Morrissey89}, which is suited
for describing the kinematics of heavy residues, and with the
prescription of Goldhaber\cite{Goldhaber74}.
%
%
%
%
%

    The production cross section of evaporation residues (in this case 
we consider any evaporation pattern where intermediate mass fragments 
are not produced by fission in the first decay steps) varies 
monotonically as a function of the mass number in agreement with 
the expectation that also the excitation energy of the hot fragments 
has a monotonic dependence (see fig.~\ref{fig1}).
    In this respect, the fact that the mass distribution of evaporation 
residues does not extend monotonically towards the lightest masses
is an additional indication that intermediate-mass fragments are not 
fed by evaporation residues.
	We may remark that this conclusion, clearly suggested by the production 
cross sections, can not be inferred from the velocity spectra. 
	There is in fact a gradual transition between the shape of the velocity 
distributions of the heaviest intermediate-mass fragments, which are 
dominated by the multifragmentation mode and which are almost gaussian, and the 
gaussian-like spectra of the evaporation residues. 
	For this reason, in lighter systems like $^{56}$Fe$+p$~\cite{Napolitani04}, 
where intermediate-mass fragments and evaporation residues do not occupy 
distinguishable regions of the production cross-section distribution, we 
could not identify the transition between the two processes from a similar 
experimental analysis.

	As a result of this section we can deduce the proportions between
the different decay modes in the total reaction cross section of the system
$^{136}$Xe$+p$.		
	In ref.~\cite{Napolitani07} a total reaction cross section of 
$1393\pm 72$~[mb] was obtained.
	We supposed that the Coulomb-hole mode corresponds essentially to 
binary splits.
	Although the production of the heavy fragments in the binary 
splits could not be disentangled, it must be equal to the
contribution of the corresponding light partners, which we measured
for the greatest part, till $A=30$, where the production cross 
section already drops to very small values. 
	Under this assumption, the portion of total cross section 
related to the Coulomb-hole mode could be deduced: we found a value 
of $59.0\pm 3.7$~[mb].
	The Weisskopf calculation of residues which evaporated protons,
neutrons and alpha particles gives an indicative estimation of the
corresponding fraction of total cross section of around 90\%.
	The contribution of the multifragmentation mode can not be deduced 
because neither the fragment multiplicity of the process, nor the 
total yield of this mode could be deduced from the measurement.
	The sum of the production cross sections related to the multifragmentation 
mode which were measured up to silicon is of $8.6\pm 0.6$~[mb].
    By subtracting the measured contribution of the Coulomb-hole mode and 
the calculated fraction of evaporation residues to the measured total 
reaction cross section we estimate an indicative cross section of 
around $35/\langle M\rangle$~[mb] for the multifragmentation mode, where 
$\langle M\rangle$ is the corresponding mean fragment multiplicity, 
supposed to be larger than two.

%
%
\begin{figure}[b]\begin{center}
\includegraphics[angle=0, width=0.6\textwidth]{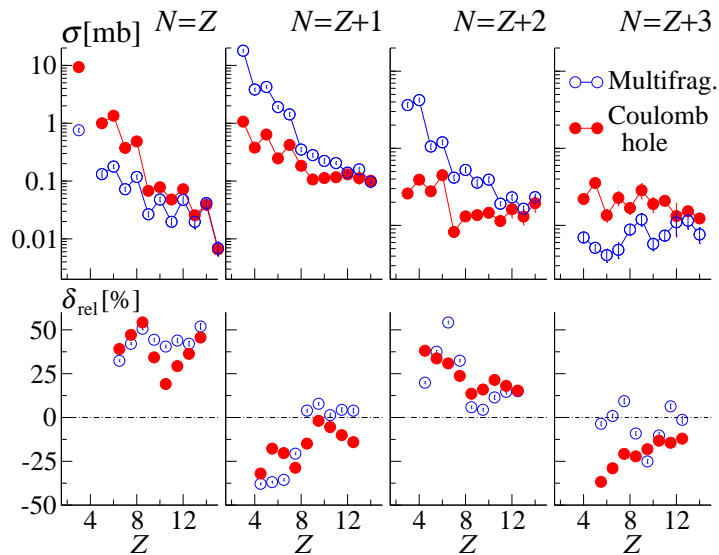}
\end{center}\caption
{
	Upper row.
	Chains of nuclide cross sections with a given value of $N-Z$
present an ``even-odd'' staggering.
	Lower row.
	Amplitude of the staggering analysed by the method of 
Tracy~\cite{Tracy72}. It is positive when the staggering is 
``even-odd'' and negative when the staggering reverses.
}
\label{fig12}
\end{figure}
\subsubsection{Fine structure in the nuclide cross sections}
	A close analysis of the nuclide cross sections reveals a 
staggering of large magnitude.
	Such a feature has been observed since long time at different 
bombarding energies (to indicate part of several contributions, 
see refs.~\cite{Sanders89,Beck92,Cavallaro98,Yang99,Winchester2001,Lombardo2011})
	The staggering in the nuclide cross sections is particularly evident 
along chains of nuclides having the same $N-Z$ value. 
	This kind of selection, which was already applied in 
refs.~\cite{Ricciardi04a, Napolitani04a} and employed for the 
system $^{136}$Xe$+p$ in ref.~\cite{Napolitani07}, is now presented 
in fig.~\ref{fig12} for the two decay modes distinctly.
	The upper row presents the measured cross sections and the 
lower row quantifies the staggering amplitude according to the 
procedure introduced by Tracy~\cite{Tracy72}.
	Despite the smoother evolution of the staggering strength for
the Coulomb-hole mode, we observe the same general phenomenology for the 
two modes.
	In a simple Weisskopf picture this staggering was described as a 
manifestation of the very last steps of the decay path, independent 
of the entrance channel; the number of excited levels of the mother 
nucleus that could decay into the daughter nucleus determines the 
probability of a channel.
	In particular, even-mass nuclides show a smooth variation of 
the separation energy as a function of the element, once shifted 
by the pairing gap.
	Therefore, their formation is enhanced due to the pairing gap, which 
determines a larger number of excited levels of the odd-mass mother
nuclei available for the decay into the ground states of even-even 
daughters rather than odd-odd daughters.
	This explains the even-odd staggering for $N=Z$ and $N=Z+2$.
On the other hand, the formation of odd-mass residues is not 
determined by the position of their ground states, which are 
ordered along the same mass parabola for one given mass number.
	Their formation reflects the structure of the separation energy.
	Since neutron-rich nuclei have lower separation energy for 
neutrons, their formation reflects the structure of the neutron 
separation energy and odd elements are favoured.
	This explains the reversal of the even-odd staggering for $N=Z+1$ 
and $N=Z+3$.
	For the same reason, the staggering becomes less pronounced in
the mass-number distribution of cross-sections
when more-neutron-rich systems are considered. 
	While chains of nuclides with negative values of $N-Z$ 
(proton-rich side of the nuclear chart, see refs.~\cite{Napolitani04}) 
contribute to an even-odd staggering, chains of nuclides with positive 
values of $N-Z$ contribute to an even-odd staggering only for even 
values of $N-Z$ (even masses) and contribute to a reversed staggering for odd 
values of $N-Z$ (odd masses).
	When dealing with more neutron-rich systems, which present a larger extension
of the nuclide production over neutron-rich isotopes, the sum of the even-odd staggering
of the even chains and the reversed staggering of the odd chains lead therefore to a 
compensation, which is visible in fig.~\ref{fig6}.
	This simple picture is sufficient to conclude from the large
amplitude of the staggering that the secondary decay affects largely
and in the same manner the production of intermediate mass fragments
from both decay modes.
	We should however remark that a quantitative description of these
effects requires first of all a more complete approach, like the 
Hauser-Feshbach formalism or one of its extensions~\cite{Matsuse04}, and the 
inclusion of Wigner terms in describing the $N=Z$ chain.
	In addition, it requires the inclusion of decays which go beyond
the Hauser-Feshbach approach for compound-nucleus decay, like
cluster emission~\cite{Beck04,vonOertzen06}. The study of such a 
process is of great relevance for describing the formation of the 
lightest fragments; for a comprehensive review on this subject see 
refs.~\cite{Clusters} 

\subsection{Connection between the entrance channel and the exit channel} \label{section5B}
%
%
\begin{figure}[b]\begin{center}
\includegraphics[angle=0, width=0.6\textwidth]{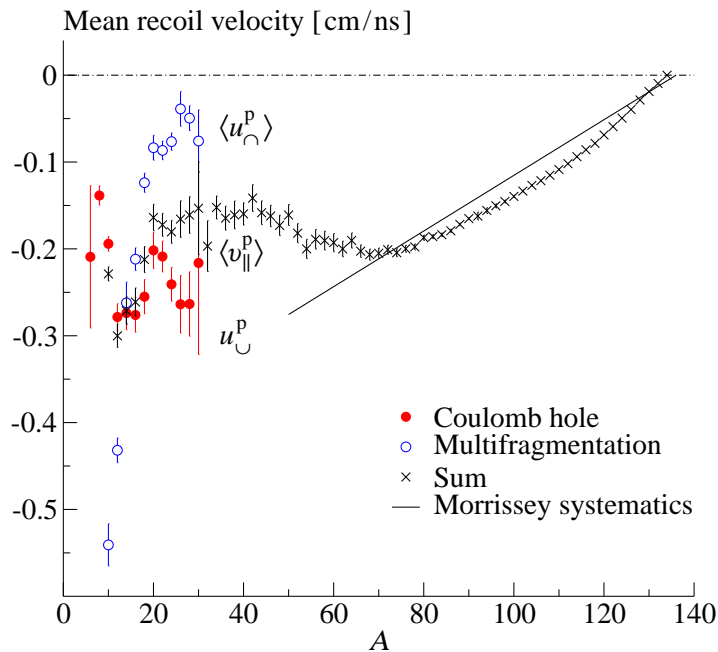}
\end{center}\caption
{
	Evolution as a function of the mass of the residue of the 
mean recoil velocity, deduced for the two kinematic modes 
separately for lower masses than $A=30$.
The mean value of the reconstructed full velocity distribution is 
shown for the whole mass range.
The systematics of Morrissey~\cite{Morrissey89} is shown for 
comparison.
}
\label{fig13}
\end{figure}
	The precise knowledge of the mean recoil velocity of the 
ending fragments of the decay process provides indications on the 
connection between the exit channel and the entrance channel.
	Such a study profits from the high-resolution measurement
of the inclusive experimental approach.
	However, this analysis can not be based on single-event 
observables, due to the large fluctuations induced by the Fermi 
momenta of abraded nucleons and the recoil due to the decay.
	In order to smear out these effects, this analysis is based 
on employing, as an inclusive quantity, the mean value of the source 
velocity of one specific component of the velocity distribution in
the projectile frame.

	It is evident that a study of the entrance channel based on 
single-event observables, concerning for instance the information 
on the excitation energy, would go beyond the specificities of this
inclusive experimental method.
	On the other hand, the aim of this study is to join the measurement
of the full production, the isotopic identification, and the high 
resolution of the kinematic observable.
\subsubsection{Mean recoil velocities}
	Fig.~\ref{fig2} shows that a systematic 
displacement in recoil velocity between the two kinematic modes,
fission and multifragmentation, could be resolved with high 
precision since the two components are measured as parts of the 
same spectrum in the same set of experimental runs. 
    The distinction of the two kinematic modes is a signature of
the interplay of two distinct exit channels; their relative 
displacement, which characterises all spectra of the 
intermediate-mass fragments shown in fig.~\ref{fig4}, attributes 
the two exit channels to different entrance channels.
	This displacement is in fact connected to the mean momentum 
transfer, which reflects the violence of the collision, and which 
evolves in different ways according to the decay mode.
    Such study is presented in fig.~\ref{fig13}, where three mass
regions can be distinguished.
\begin{enumerate}
\item Heavier masses than half the mass of the projectile ($A>68$). 
	They are necessarily the heaviest fragments produced in the 
reaction. 
	They cannot be the lighter fragments of a binary decay, and 
they are not typical multifragmentation products, which should be 
more than two and should have comparable size. 
\item Intermediate-mass fragments with $A\le 30$.
	Explicit information on the source velocities of the two 
kinematic modes could be obtained experimentally.
\item Fragments in the range $30<A\le68$. This is in the
minimum of the production as shown in fig.~\ref{fig11}; only the 
average source velocity could be measured, without any kinematic 
distinction of the decay process.
\end{enumerate}

	The region of heavier mass numbers than half the mass of
the projectile is dominated by the presence of evaporation 
residues.
	Here, the mean recoil $v_{\|}^\beam$ evolves almost 
linearly with the mass loss from the projectile velocity down 
to around $-0.2$ cm/ns with respect to the projectile frame.
	The intermediate-mass production attributed to the 
Coulomb-hole mode should originate from the decay of heavy 
residues which evaporate not only nucleons but also heavier 
fragments. 
	In the process, the accompanying heavy residues should have 
suffered a large loss of mass.
	Therefore, it is reasonable that the largest negative mean 
source velocities, of around $-0.25$ cm/ns, reflected by 
intermediate-mass fragments issued of binary splits, are close to 
the source velocities which characterise residues of around half 
the projectile mass ($A\sim68$).
	With less statistical significance, the data on	
intermediate-mass fragments from the coulomb-hole mode suggest
also the tendency of reducing the mean recoil with increasing mass: 
the mean recoil varies from around $-0.2$ cm/ns for $A\sim5$ to 
around $-0.25$ cm/ns for $A\sim30$, in the projectile frame. 
	This is reasonable, because the emission of heavier 
intermediate-mass fragments from binary decay is expected for
higher excitation energies; at the same time, higher excitation 
energies correspond to smaller impact parameters, reflected by
less heavy partners which also tend to have larger negative mean 
velocities.

	Altogether, the experimental results on binary-decay events
suggest that the source velocities can be well explained by the 
empirical systematics of Morrissey~\cite{Morrissey89}: both in the 
heavy ($A>68$) and in the light (Coulomb-hole mode measured up to 
$A=30$) residues, there is a tendency towards more violent 
collisions from the heaviest, respectively lightest, masses towards 
more symmetry.
	In particular, in the case of binary decays, three 
quantities are correlated: the evolution of the system towards more 
symmetric masses is a signal of increasing friction for less 
peripheral impact parameters, and is reflected in the mean-source 
velocity.
	In the case of light intermediate-mass fragments issued of 
multifragmentation, the correlation between two of these 
quantities, the size and the impact parameter, is also well known 
from previous experiments~\cite{Schuttauf96}, where the fragment 
size was observed to decrease with increasing energy deposition, 
for more central collisions.
%

	The mean recoil of the intermediate-mass fragments issued
of multifragmentation could be measured in a limited mass range 
($A\le 30$). 
	For the heaviest fragments, it approaches almost the 
velocity of the projectile around $A=30$.
	For the lighter fragments, we find a strong tendency to 
larger negative values in the projectile frame.
	Both the velocities of multifragmentation products
approaching zero
around $A=30$ and the strong trend to very large negative 
values towards $A\sim10$ have no reasonable counterpart in the 
heavy residues.
	This is not surprising, since, on the one hand, the heavy 
fragments cannot be produced in a typical multifragmentation 
process; on the other hand, also the convex pattern in the 
distribution of invariant cross section of the multifragmentation 
products contradicts the simultaneous emission of a heavy partner.
	For larger masses than around $A=15$, the source velocities 
of the intermediate-mass fragments from multifragmentation variate 
in the same range of the source velocities of both the heavier and 
the lighter partners of binary splits; still, the two decay modes 
can be distinguished by the shape of the velocity distribution.

\subsubsection{The mean recoil velocities as a possible indicator of the impact parameter.}
%
%
%
	The new additional information of this experiment, that the 
size of the intermediate-mass fragments correlates with the mean 
recoil in the whole mass range and for both the two kinematic modes,
makes plausible to establish a further correlation between 
the mean source velocity and the impact parameter also for the 
multifragmentation products.  
	We think this is an important generalisation of the 
interpretation of the kinematic information, which combines results 
from previous experiments with the new findings of this work.
	In this respect, this experiment extends the well known 
Morrissey kinematics which was established for masses very close to 
the heavy reaction partner, all the way down to the 
intermediate-mass fragment region. 
	Using the well established correlation of the fragment size 
in the multifragmentation regime with the impact parameter, we show 
that the fragment size also correlates with the source velocity. 
	This leads to the conclusion that the source velocity
is another variable suitable to characterise the impact parameter.

	However, differently from the case of the heavy residues, 
the mean source velocity tagged by the intermediate-mass 
fragments produced in multifragmentation may manifest a reversed 
dependence, reducing to very small absolute values in the projectile frame.
    The experimental disentangling of the two kinetic modes could 
be extended up to rather heavy fragments: it allowed to make 
evident the presence of a reversal point of such dependence in the 
region of the heaviest intermediate mass fragments, around $A=25$, 
determined by the multifragmentation products.
    Such finding results from the high-resolution of the inclusive 
method.
    For heavier masses, the measurement could no more distinguish 
the two kinetic modes.
    However, the reversal is also evident, even if smoothed, by 
analysing the overall behaviour of the sum of the two kinetic modes, 
which could be measured for the full range of masses; such 
observable connects the evolution of the intermediate-mass 
fragments with the behaviour of the heavy residues.    
	Due to this effect, the impact parameter is not uniquely 
defined by the mean source velocity in the projectile frame for any 
fragment size.
    This ambiguity can be controlled by additionally accounting 
for the information on the size of the largest fragment: the 
products which appear with the same mean velocity in the projectile 
frame are very much separated, if we consider the mass of the 
heaviest fragment formed in the reaction as an additional 
criterion.
    In this respect, a correlation of the mean velocity in the 
projectile frame with the impact parameter could be established 
also in the intermediate-mass region.

We may stress that the relation between energy loss and momentum 
transfer probed by the multifragmentation data should not be regarded 
within a two-body picture which connects the amount of dissipated 
energy with the momentum transfer like in dissipative two-body collisions. 
	In this case there would appear a contradiction: multifragmentation 
products should gain the highest excitation energies, but we 
observe that the absolute value of their momentum transfer reaches 
the lowest values. 
	Indeed, this is possible when more than two objects are present 
in the exit channel.
	In our experiment there are two objects in the entrance channel, 
the target proton and the Xe projectile, but many more particles 
may appear in the exit channel; in this respect, it is possible to fulfil 
energy and momentum conservation even in cases where the velocity of 
the Xe-like fragment is almost unchanged and, at the same time, its 
excitation energy is very high. Such a condition would be associated 
to an asymmetric emission of cascade or pre-equilibrium particles. 
	Of course, an exclusive experiment could reveal this configuration 
directly, but at present an inclusive approach has the advantage of giving
a more precise measurement of the central region of the longitudinal 
velocity distribution connected to small angles.
	Among others, one experimental proof which shows clearly 
that a many-body configuration in the exit channel can not be described 
as a two-body kinematics is given by the signature studied in the 
work~\cite{Ricciardi03}, where intermediate-mass fragments produced in 
relativistic heavy-ion collisions are predominantly forward emitted.
	When spallation reactions lead to multifragmentation, the scenario
does not change substantially with respect to the heavy-ion collisions. 
	In both cases, a picture based on two-body kinematics, dominated by 
the collision term, and in absence of collective effects, is incomplete.
	The difference is in the experimental signatures, which are more 
pronounced in heavy-ion collisions than in spallation.
	In particular, when proton-induced reactions produce some collective 
effects, like some thermal pressure created in the cone around the proton 
traversing the nucleus, these effects are still small in the 1 GeV 
regime~\cite{Beaulieu00}.

%
\subsubsection{Possible traces of phase transitions?}
	Supposing that in proton-induced reactions at high 
incident energies dynamical effects are negligible, we find that 
the dependence of the recoil observable as a function of the mass 
number is very different for the Coulomb-hole and 
multifragmentation modes.
	As depicted in fig.~\ref{fig13}, for the first mode it 
varies in a reduced range, while for the second it presents a 
strong dependence.

	Keeping in mind that the recoil observable can be 
interpreted as an indicator of the impact parameter, this behaviour 
may be interpreted tentatively as follows.
	The Coulomb-hole mode is associated to a small range 
(with respect to the uncertainties) of recoil velocities 
in the projectile frame: this may indicate that, within the picture 
of a Liquid-Fission phase transition, this mode is also associated 
to small energy fluctuations when the system is situated in the 
coexistence region, and it corresponds to a phase transition with 
vanishing latent heat, as expected for the Liquid-Fission phase 
transition~\cite{Lehaut09, Gulminelli09,Gulminelli03,Chaudhuri09}.
	The situation changes completely for the multifragmentation
mode, which is associated to a very large range of recoil 
velocities in the projectile frame; this may correspond to a strong 
dependence of the fragment configuration on the excitation energy, 
which corresponds to a finite latent heat.
	If we consider the heavy residues and the intermediate-mass
fragments issued of multifragmentation, and we exclude the 
Coulomb-hole mode, the plot of fig.~\ref{fig13} gives a good 
information on the evolution of the recoil as a function of the 
heaviest fragment.
	With increasing absolute values of the average recoil in 
the projectile frame, the system firstly explores a pattern 
characterised by only heavy evaporation residues, close to the mass 
of the projectile; it then explores a configuration where both the 
intermediate-mass-fragment production and the residues coexist; 
finally, it evolves towards a situation where the very large 
average absolute recoil velocity in the projectile frame 
corresponds only to the multifragmentation mode and no more to the 
residue production.
	These inclusive observables are not incompatible with 
signals of bimodality, which were widely investigated in recent 
exclusive experiments~\cite{Tabacaru03,Lopez06,Pichon06,Bonnet07,Bonnet07b,Bonnet08,Bruno08},
and which are proposed as a manifestation of the liquid-gas phase 
transition in nuclear systems.
	Our interpretation, based on average observables, follows 
closely this line, but it should be corroborated and quantified by 
exclusive experiments.
%
%
\section{
    Conclusions	and prospects				\label{section6}
}
	In this report we developed a specific procedure for the 
deconvolution of the distributions of longitudinal velocities of 
intermediate-mass fragments, which were measured inclusively and
identified in atomic and mass number.
	Technically, the experimental approach we adopted in 
measuring the characteristics of the intermediate-mass fragments 
has several unique advantages: it avoids the contamination from 
target-like reaction partners with low kinetic energy which is a 
difficulty in direct kinematics experiments, and the overloading by 
the primary beam at small forward angles which challenge 
inverse-kinematics experiments in absence of a spectrometer; it 
also profits from the absence of the low-energy detection threshold 
which characterises silicon detectors, and profits from the absence 
of magnetic aberrations of the achromatic setting of the 
small-acceptance spectrometer.
	As a main disadvantage, the reduced acceptance precludes to 
measure particle correlations.

	As a mayor general result, we could disentangle and analyse 
quantitatively, in terms of invariant cross sections and in terms 
of production cross sections two kinematic modes for the exit 
channel in the spallation of $^{136}$Xe at 1 GeV per nucleon,
for several isotopes of light elements ranging from lithium to
silicon.

	One mode is produced by a Coulomb shell in velocity space 
and results in a convex shape in the distribution of invariant cross section,
related to events with low fragment multiplicity and large 
asymmetry in mass number.
	This mode is attributed to asymmetric binary decays, and 
the fission velocity deduced from the measurement is consistent 
with the binary split of a system close to the mass of the 
projectile; this is however less true for elements lighter than carbon, which exhibit
very high velocities, incompatible with standard calculations.
	The results are in line with previous similar studies of the formation
of intermediate-mass fragments in the lighter system
$^{56}$Fe$+p$~\cite{Napolitani04}
and in the heavier system $^{238}$U$+p$~\cite{Ricciardi06}
at the same incident energy of 1 GeV per nucleon.

	Another mode reflects in a convex shape in the distribution
of invariant cross section, related to events with large fragment 
multiplicity and small asymmetry in mass number.
	This mode is attributed to multifragmentation. 
	As a relevant result of the present analysis we indicate that, 
despite the small fraction of the total cross section which feeds 
multifragmentation, the contribution of this process to single 
nuclide cross sections is found to be large for intermediate-mass 
fragments and even dominant for some specific nuclides.
	Such a large contribution and, more generally, the large 
production of light elements, may have consequences for 
technological applications of spallation, such as transmutation and
safety.

%
	The main result of this work is the identification of these 
two decay processes, and the extraction of kinematic and production 
observables related to the exit channel and associated to the two 
processes separately.
	We measured the characteristics of the ending fragments of 
the decay process, which give valuable information on the heaviest 
reaction product; the latter is even a fundamental probe of the
thermodynamic properties of finite nuclear systems.
	we also concentrated on the sequence of evaporation decays, 
reflected in the staggering of nuclide cross sections; such 
staggering has a comparable amplitude for the two decay 
modes, and we attributed it to a large fraction of secondary decay; 
unfortunately, such process has the effect of smearing out the 
observables which could connect the entrance channel to the exit 
channel.

	However, signals of this connection are found back in the 
dependence of the mean recoil as a function of the mass number.
	By combining results from previous experiments with the new
information of this work, we established the inclusive mean source 
velocity as a valuable indicator of the impact parameter.
	By relying on such indicator, we firstly confirmed that the 
Coulomb-hole and the multifragmentation modes, reflect different 
processes.
	Then, on the more speculative side, we placed our study in 
the context of the recent experimental and theoretical 
investigations on the phase transitions in nuclear systems.
	We advanced the idea that the two modes could also be 
related to the phase-transition phenomenology in the finite nuclear 
system.
	The Coulomb-hole mode exhibits compatible features with a 
Liquid-Fission phase transition, characterised by vanishing latent 
heat.
	The characteristics which describe the multifragmentation 
mode are in qualitative accordance with recent observations of a 
bimodal distributions of the largest fragment charge: for the less 
violent collisions the multifragmentation mode does not appear and 
only the heaviest evaporation residues are produced.
	The multifragmentation mode then gradually prevails for
more and more violent collisions till it dominates the decay 
pattern for the most violent collisions, leading to the production 
of the smallest residues.
	This behaviour was found in the study of projectile-like 
fragments formed in (se\-mi-)\-pe\-ri\-ph\-eral collisions of heavy 
ions at Fermi energies and it was based on exclusive event 
selections; it was interpreted as a robust signal of the Liquid-Gas 
phase transition in the finite nuclear 
system~\cite{Tabacaru03,Lopez06,Pichon06,Bonnet07,Bonnet07b,Bonnet08,Bruno08}.
	In this case, the dense phase was identified with residue-like
fragments, after the suppression of fission events.

	Relativistic nuclear reactions induced by hadrons were
already employed for investigating the statistical nature of the 
decay process~\cite{ISIS}.
	The fraction of incident energy which is transferred to the 
prefragment appears almost completely as single-particle excitation, 
while collective effects like compression are relatively small;
for this reason, we suggest that spallation 
reactions may be also suited for distinguishing the signals of 
Liquid-Fission and the signals of the Liquid-Gas phase transitions 
in finite nuclear systems.
	In the same context, we advance the view that a method of 
inclusive event selection like the one presented in this paper and, 
more generally, event selections based on kinematic observables, 
could be relevant in a further generation of experiments also at 
Fermi energies, where the measurement of isotopic and kinematic 
observables in correlation with light-fragments is a mayor request 
and a foreseen development of the previous-generation exclusive
experiments.
	This is already the direction followed by very recent 
experiments, where a spectrometer is coupled to a $4\pi$ 
detector array~\cite{Chbihi07}.
	In this respect, high-precision momentum measurements and 
the nuclide identification of the residues, from inclusive 
approaches, can offer a further tool of event selection if coupled 
to correlation measurements, from exclusive approaches.

%
\appendix
\section{
    Velocity reconstruction with multiple sources   \label{appendixA}
}
%
	In this section we give a general description of the
deconvolution of the measured longitudinal velocity 
distributions shown in fig.~\ref{fig2}, which leads to the
invariant cross sections shown in fig.~\ref{fig4}.
	This procedure is based on the three conditions listed
in section \ref{section3}.
	For generality, in this section we suppose that one 
velocity spectrum results from a continuous distribution 
of sources along the beam axis.

	The deconvolution imposes to change among three 
reference frames, the laboratory frame, the projectile 
frame, and the reference frame of one given isotropic source, 
labelled L,P,u, respectively.
	We define the velocity vectors $\vvec^{\beam}$, $\uvec^{\beam}$, and
$\vvec^u = (\vvec-\uvec)$, to indicate the velocity of 
fragments in the projectile frame, the velocity of a given 
source $u$ in the projectile frame, and the velocity of 
fragments in the frame of a given source $u$, respectively

	We intend to pass by deconvolution from the measured
longitudinal velocity distribution $\dyield / \diff\vpar^{\beam}$, which depends on
the acceptance, to the (non-relativistic) invariant cross section
$\si(\vvec^\beam)$, which is not depending on the acceptance.
	Since we must proceed through unfolding over the distribution
of sources, we define the corresponding distributions associated to
a given source $u$ as in the following scheme:
	\begin{center}
	\begin{tabular}{l l l}
	              & full space & source $u$
	\vspace{5pt}\\ 
	acceptance    $\,\,\,$ & $\dyield / \diff\vpar^{\beam} \,\,\,\,\,\,$ & $\diff[\delta_u I(\vpar^u)] / \diff\vpar^u \,\,\,$
	\vspace{5pt}\\ 
	no acceptance $\,\,\,$ & $\si(\vvec^\beam)$ & $\deltausigmaI(\vvec^u)$
	\end{tabular}
	\end{center}
$\diff[\delta_u I(\vpar^u)] / \diff\vpar^u$, where
$\delta_u$ stands for $\diff / \diff \uvec^{\beam}$, is the portion
of the measured longitudinal velocity distribution which we 
attribute to a given source $u$, and 
$\deltausigmaI(\vvec^u)$ is the portion of the 
invariant cross section which we attribute to the source $u$.

    The measured production yield $\dyield / \diff\vpar^{\beam}$ of one individual 
nuclide is the integration over the emitting sources of the quantity 
$\delta_u I(\vpar^\beam)$, which is the portion of the spectrum
associated to the sources lying in the interval 
$[u^\beam,u^\beam+\diff u^\beam]$:
\begin{equation}
	\yield  =
		\int\limits_{u^\beam}
		\frac{\diff}{\diff\vpar^{\beam}}
		\delta_u I (\vpar^\beam)\,
		\diff u^\beam
	.
\label{eqA1}\end{equation}
	The angular acceptance of the magnetic spectrometer, with 
an almost circular aperture of
$\aver{\alpha^\Lab(\varphi)}\approx 15$~mr in the laboratory frame 
(the azimuthal angle $\varphi$ around the beam axis is introduced 
to take into account the deviation of the acceptance shape from a 
circle) imposes that a limited selection of the emission-velocity 
space could be explored and only the selected part of $\delta_u I$ 
appears in the measured yields.
	The Lorentz transformation attributes to this selected portion 
a conical boundary along the beam axis in the frame of one emitting 
source, defined by the velocity vector 
$\vvperm^u$, of magnitude 
$\vperm^u = \gamma_{\|}^\Lab(u^\Lab+\vpar^u)\mathrm{tg}\alpha^{\Lab}(\varphi)$,
and orientation along the azimuthal angle $\varphi$ in the plane
orthogonal to the beam axis; $\vpar^u$ is the longitudinal
projection of the boost $\vvec^u$ in the frame of the source $u$,
imparted to the emitted fragment.
	By moving to the frame of the source $u$, the contribution 
$\diff[\delta_u I(\vpar^u)]/\diff\vpar^u$ of the source $u$ to the 
measured yields is described by the integration of the transverse 
velocity $\vper^u$ over the portion of the plane selected by the 
ensemble of vectors $\vvperm^u$, so that
\begin{equation}
	\frac{\diff[\delta_u I(\vpar^u)]}
	{\diff\vpar^u}
		=
		\int\limits_{0}^{2\pi}
		\int\limits_{0}^{|\vperm^u|}
		\vper^u\,
		\ddVu{[\delta_u \sigma(\vvec^u)]}\,
		\diff\vper^u
		\diff\varphi\,
	,
\label{eqA2}\end{equation}
	where $\sigma$ is the production cross section for the nuclide
associated to the velocity spectrum.
	Since in our case the velocities in the source frame are
not relativistic, the argument of the integral in eq.~(\ref{eqA2})
is equivalent to the invariant 
cross section as a function of the velocity in the source frame:
\begin{equation}
	\frac{\diff[\delta_u\sigma(\vvec^u)]}{\diff\vvec^u}
		=
    \frac{\mathrm{m}^2}{\mathrm{c}^2}		
    E^u   
    \frac{\diff[\delta_u \sigma (\vvec^u)]}{\diff\pvec^u}
		=
    \frac{\mathrm{m}^2}{\mathrm{c}^2}\deltausigmaI(\vvec^u)	\,	
	,
\label{eqA3}\end{equation}
where $E^u$ and $\pvec^u$ indicate the total energy and the momentum in 
the source frame, respectively.
	Introducing the invariant cross section in the integral we write:
\begin{equation}
	\frac{\diff[\delta_u I(\vpar^u)]}
	{\diff\vpar^u}
		=
		\frac{\mathrm{m}^2}{\mathrm{c}^2}
		\int\limits_{0}^{2\pi}
		\int\limits_{0}^{|\vperm^u|}
		\deltausigmaI(\vvec^u)\,
		\vper^u\diff\vper^u\,
		\diff\varphi\,
	,
\label{eqA4}\end{equation}
	This expression does not impose any constraint for the 
distribution of emitted fragments.
	Under the assumption of isotropy 
$\deltausigmaI(\vvec^u) = \deltausigmaI(v^u) = [1/4\pi(v^u)^2]\diff[\delta_u\sigma(v^u)] / \diff v^u$
and by substituting the 
integration variable $\vper^u$ with 
$v^u = \sqrt{(\vpar^u)^2 + (\vper^u)^2}$ we can
write (indicating $\vvpar^u$ the vector of magnitude $\vpar^u$,
oriented along the beam direction):
\begin{equation}
	\frac{\diff I(\vpar^\beam)}
	{\diff\vpar^\beam}
		=
		\frac{\mathrm{m}^2}{\mathrm{c}^2}
		\int\limits_{u^\beam}
		\int\limits_{0}^{2\pi}
		\int\limits_{|\vpar^u|}^{|v^u|}
		\deltausigmaI(v^u)\,
		v^u\diff v^u\,
		\diff\varphi\,
        \diff u^{\beam}
	,
\label{eqA5}\end{equation}
where the integral boundaries are
$|\vpar^u| = |\vpar^{\beam} - u^{\beam}|$ and
$|v^u| = \sqrt{(\vpar^{\beam}-u^{\beam})^2+(\vperm^u)^2}$.

	The distribution of invariant cross section $\si (v^u)$,
presented in fig.~\ref{fig4} was deduced by the deconvolution of the 
integral eq.~(\ref{eqA5}).
    The deconvolution can be solved only by restricting to a finite
number of degrees of freedom: this implies that the number of sources
is limited (the integral over $u^{\beam}$ is converted into a discrete
sum over the sources).
	With this general prescription we constructed an 
axial-symmetric distribution of invariant cross section in the beam 
frame $\si(\vpar ^\beam)$ as the superposition of isotropic 
emission patterns, each one defined by one individual source:
\begin{equation}
	\frac{\mathrm{c}^2}{\mathrm{m}^2}
    \frac{\diff \sigma}{\diff \vvec^{\beam}}  =
    \si(\vvec^\beam)  =
	\int\limits_{u^\beam}
	\delta_u \sigma_{\mathrm{I}u}(|\vvec^{\beam} - \uvec^{\beam}|)\,
	\diff u^\beam \,
	.
\label{eqA6}\end{equation}
%
%
	This spectrum would be obtained directly from the experimental
measurement only if two conditions were satisfied. First, the angular 
acceptance should not impose any cut. Second, also the transversal 
components of the velocity should be measured.
%
%
%
\section{
    Parameters of the invariant cross sections  \label{appendixB}
}
    The analysis described in this work results from the deconvolution
procedure introduced in section~\ref{section3} and applied to the
analysis of section~\ref{section4}, so as to produce the results
sketched in fig.~\ref{fig3}.
    As an outcome, the procedure gave the parameters listed in 
table~\ref{tab3}.
%
%
\begin{table}[bh!]
\caption
{
\label{tab3}
	Parameters of the invariant cross sections, which determine the 
plots in fig.~\ref{fig4}, fig.~\ref{fig10} and fig.~\ref{fig13}.
	The parameters are 
the mean recoil of the Coulomb-hole mode 
	$u_{\smallsmile}^\beam$,
the ridge of the ring associated to the Coulomb-hole mode in the invariant cross section distribution 
	$v_{\mathrm{peak}}$,
the standard deviation of the Coulomb-hole mode in the invariant cross section distribution 
	$\sigma_{\| \smallsmile}^\beam$,
the mean recoil of the multifragmentation mode 
	$\langle u_{\smallfrown}^\beam\rangle$,
and the average standard deviation of the components of the multifragmentation mode in the invariant cross section distribution 
	$\langle \sigma_{\| \smallfrown}^\beam\rangle$.
}
\begin{tabular}{l}
$\quad A $
$\qquad\; u_{\smallsmile}^\beam$
$\qquad\quad\; v_{\mathrm{peak}}$
$\qquad\;\; \sigma_{\| \smallsmile}^\beam$
$\qquad\quad \langle u_{\smallfrown}^\beam\rangle$
$\quad\;\; \langle \sigma_{\| \smallfrown}^\beam\rangle$
\vspace{3pt}
\\
\hline
\vspace{-3pt}
\\
\begin{tabular}{l l l l l}
$\qquad$   Li \\
$\,\,\,\;\;6\;\; -0.21(11)   $&$ 2.49(21)  $&$ 0.497(43) $&$ -1.04(14)  $&$ 0.90(12)  $\\
$\,\,\,\;\;7\;\; -0.21(11)   $&$ 2.37(21)  $&$ 0.471(43) $&$ -1.26(14)  $&$ 0.87(12)  $\\
$\,\,\,\;\;8\;\; -0.091(21)  $&$ 2.339(40) $&$ 0.433(45) $&$ -0.749(62) $&$ 0.930(23) $\\
$\qquad$   Be \\
$\,\,\,\;\;9\;\; -0.1839(74) $&$ 2.292(14) $&$ 0.448(11) $&$ -0.666(21) $&$ 0.836(21) $\\
$\,\,\,   10\;\; -0.1018(99) $&$ 2.170(18) $&$ 0.454(18) $&$ -0.636(21) $&$ 0.850(21) $\\
$\,\,\,   11\;\; -0.052(21)  $&$ 2.083(39) $&$ 0.366(42) $&$ -0.329(73) $&$ 1.286(32) $\\
$\,\,\,   12\;\; -0.024(28)  $&$ 2.107(51) $&$ 0.427(44) $&$ -0.270(32) $&$ 0.826(21) $\\
$\qquad$   B \\
$\,\,\,   10\;\; -0.389(23)  $&$ 2.270(43) $&$ 0.498(38) $&$ -0.25(13)  $&$ 0.890(22) $\\
$\,\,\,   11\;\; -0.2405(98) $&$ 2.180(18) $&$ 0.462(14) $&$ -0.564(21) $&$ 0.821(21) $\\
$\,\,\,   12\;\; -0.316(23)  $&$ 1.985(43) $&$ 0.478(32) $&$ -0.485(23) $&$ 0.918(23) $\\
$\,\,\,   13\;\; -0.245(80)  $&$ 2.10(15)  $&$ 0.45(13)  $&$ -0.567(18) $&$ 0.702(18) $\\
$\qquad$   C \\
$\,\,\,   11\;\; -0.372(36)  $&$ 2.361(67) $&$ 0.579(87) $&$ -0.571(50) $&$ 1.07(11)  $\\
$\,\,\,   12\;\; -0.315(12)  $&$ 2.108(21) $&$ 0.457(17) $&$ -0.438(18) $&$ 0.682(17) $\\
$\,\,\,   13\;\; -0.256(15)  $&$ 2.052(28) $&$ 0.451(21) $&$ -0.352(16) $&$ 0.633(16) $\\
$\,\,\,   14\;\; -0.310(18)  $&$ 1.944(34) $&$ 0.424(24) $&$ -0.294(25) $&$ 0.984(25) $\\
$\,\,\,   15\;\; -0.190(76)  $&$ 1.98(14)  $&$ 0.41(13)  $&$ -0.325(87) $&$ 0.850(78) $\\
$\qquad$   N \\
$\,\,\,   13\;\; -0.187(55)  $&$ 2.17(10)  $&$ 0.457(95) $&$ -0.291(19) $&$ 0.749(19) $\\
$\,\,\,   14\;\; -0.297(29)  $&$ 2.063(53) $&$ 0.470(72) $&$ -0.501(27) $&$ 0.631(37) $\\
$\,\,\,   15\;\; -0.248(26)  $&$ 1.997(47) $&$ 0.396(30) $&$ -0.181(37) $&$ 0.680(22) $\\
$\,\,\,   16\;\; -0.257(29)  $&$ 1.908(53) $&$ 0.384(45) $&$ -0.055(14) $&$ 0.546(14) $\\
$\,\,\,   17\;\; -0.177(74)  $&$ 1.76(13)  $&$ 0.424(83) $&$ -0.274(12) $&$ 0.470(21) $\\
\end{tabular}
\\
\end{tabular}
\end{table}
%
%
\begin{table}[h!]
\begin{tabular}{l}
$\quad A $
$\qquad\; u_{\smallsmile}^\beam$
$\qquad\quad\; v_{\mathrm{peak}}$
$\qquad\;\; \sigma_{\| \smallsmile}^\beam$
$\qquad\quad \langle u_{\smallfrown}^\beam\rangle$
$\quad\;\; \langle \sigma_{\| \smallfrown}^\beam\rangle$
\vspace{3pt}\\
\hline\vspace{-3pt}\\
\begin{tabular}{l l l l l}
$\qquad$   O \\
$\,\,\,   15\;\; -0.16(13)   $&$ 2.06(27)  $&$ 0.62(21)  $&$ -0.235(33) $&$ 0.604(74) $\\
$\,\,\,   16\;\; -0.342(18)  $&$ 1.921(32) $&$ 0.425(33) $&$ -0.359(15) $&$ 0.604(15) $\\
$\,\,\,   17\;\; -0.273(11)  $&$ 1.913(19) $&$ 0.373(23) $&$ -0.171(21) $&$ 0.845(21) $\\
$\,\,\,   18\;\; -0.241(17)  $&$ 1.859(31) $&$ 0.366(24) $&$ -0.130(13) $&$ 0.518(13) $\\
$\,\,\,   19\;\; -0.290(43)  $&$ 1.829(79) $&$ 0.384(54) $&$ -0.165(19) $&$ 0.729(18) $\\
$\qquad$   F \\
$\,\,\,   17\;\; -0.204(62)  $&$ 2.04(11)  $&$ 0.265(93) $&$ -0.222(80) $&$ 0.917(44) $\\
$\,\,\,   18\;\; -0.329(22)  $&$ 1.843(40) $&$ 0.382(40) $&$ -0.161(16) $&$ 0.633(16) $\\
$\,\,\,   19\;\; -0.240(36)  $&$ 1.881(66) $&$ 0.351(66) $&$ -0.072(14) $&$ 0.566(17) $\\
$\,\,\,   20\;\; -0.262(20)  $&$ 1.805(37) $&$ 0.338(35) $&$ -0.014(14) $&$ 0.562(14) $\\
$\,\,\,   21\;\; -0.125(47)  $&$ 1.779(96) $&$ 0.413(68) $&$ -0.127(32) $&$ 0.631(45) $\\
$\qquad$   Ne \\
$\,\,\,   19\;\; -0.373(95)  $&$ 1.87(17)  $&$ 0.37(15)  $&$ -0.189(84) $&$ 0.578(60) $\\
$\,\,\,   20\;\; -0.222(31)  $&$ 1.853(56) $&$ 0.353(63) $&$ -0.220(15) $&$ 0.612(15) $\\
$\,\,\,   21\;\; -0.196(16)  $&$ 1.831(30) $&$ 0.360(35) $&$ -0.064(13) $&$ 0.532(13) $\\
$\,\,\,   22\;\; -0.159(17)  $&$ 1.788(30) $&$ 0.339(24) $&$ -0.021(12) $&$ 0.460(12) $\\
$\,\,\,   23\;\; -0.246(55)  $&$ 1.688(96) $&$ 0.367(75) $&$ -0.124(27) $&$ 0.498(18) $\\
$\qquad$   Na \\
$\,\,\,   22\;\; -0.231(30)  $&$ 1.782(55) $&$ 0.373(54) $&$ -0.119(12) $&$ 0.479(12) $\\
$\,\,\,   23\;\; -0.2661(83) $&$ 1.730(15) $&$ 0.330(19) $&$ -0.144(12) $&$ 0.486(12) $\\
$\,\,\,   24\;\; -0.240(10)  $&$ 1.712(18) $&$ 0.315(21) $&$ -0.081(12) $&$ 0.495(12) $\\
$\,\,\,   25\;\; -0.186(32)  $&$ 1.647(58) $&$ 0.330(42) $&$ -0.083(10) $&$ 0.400(12) $\\
$\qquad$   Mg \\
$\,\,\,   23\;\; -0.14(10)   $&$ 1.893(69) $&$ 0.33(14)  $&$ -0.141(69) $&$ 0.553(28) $\\
$\,\,\,   24\;\; -0.245(56)  $&$ 1.675(96) $&$ 0.387(87) $&$ -0.029(14) $&$ 0.567(14) $\\
$\,\,\,   25\;\; -0.323(20)  $&$ 1.587(34) $&$ 0.342(36) $&$ -0.086(17) $&$ 0.666(17) $\\
$\,\,\,   26\;\; -0.242(21)  $&$ 1.611(37) $&$ 0.342(44) $&$ -0.002(14) $&$ 0.556(14) $\\
$\,\,\,   27\;\; -0.316(79)  $&$ 1.51(13)  $&$ 0.31(11)  $&$ -0.108(54) $&$ 0.690(37) $\\
$\qquad$   Al \\
$\,\,\,   25\;\; -0.18(78)   $&$ 1.61(14)  $&$ 0.39(19)  $&$  0.03(31)  $&$ 0.39(98)  $\\
$\,\,\,   26\;\; -0.226(71)  $&$ 1.60(12)  $&$ 0.367(91) $&$ -0.079(20) $&$ 0.610(22) $\\
$\,\,\,   27\;\; -0.258(41)  $&$ 1.538(68) $&$ 0.379(83) $&$ -0.017(15) $&$ 0.586(15) $\\
$\,\,\,   28\;\; -0.284(24)  $&$ 1.542(41) $&$ 0.332(31) $&$ -0.045(15) $&$ 0.598(15) $\\
$\,\,\,   29\;\; -0.194(56)  $&$ 1.497(94) $&$ 0.31(14)  $&$ -0.073(28) $&$ 0.575(38) $\\
$\,\,\,   30\;\; -0.330(98)  $&$ 1.47(16)  $&$ 0.30(10)  $&$ -0.08(16)  $&$ 0.667(85) $\\
\end{tabular}
\\
\end{tabular}
\end{table}
%
%
\begin{table}[h!]
\begin{tabular}{l}
$\quad A $
$\qquad\; u_{\smallsmile}^\beam$
$\qquad\quad\; v_{\mathrm{peak}}$
$\qquad\;\; \sigma_{\| \smallsmile}^\beam$
$\qquad\quad \langle u_{\smallfrown}^\beam\rangle$
$\quad\;\; \langle \sigma_{\| \smallfrown}^\beam\rangle$
\vspace{3pt}\\
\hline\vspace{-3pt}\\
\begin{tabular}{l l l l l}
$\qquad$   Si \\
$\,\,\,   28\;\; -0.274(38)  $&$ 1.493(63) $&$ 0.374(60) $&$ -0.086(16) $&$ 0.659(16) $\\
$\,\,\,   29\;\; -0.333(75)  $&$ 1.42(12)  $&$ 0.373(75) $&$ -0.014(16) $&$ 0.633(16) $\\
$\,\,\,   30\;\; -0.19(16)   $&$ 1.56(26)  $&$ 0.39(19)  $&$ -0.076(16) $&$ 0.632(81) $\\
$\,\,\,   31\;\; -0.213(33)  $&$ 1.460(54) $&$ 0.295(55) $&$ -0.077(11) $&$ 0.442(11) $\\
$\,\,\,   32\;\; -0.22(17)   $&$ 1.42(27)  $&$ 0.31(15)  $&$ -0.08(24)  $&$ 0.61(25)  $\\
\end{tabular}
\\
\end{tabular}
\end{table}
$\;$
\newpage $\;$
%
%
%
%

\end{document}